\documentclass[10pt]{revtex4-1}
\usepackage{amssymb}
\usepackage{latexsym}
\usepackage{epsfig}
\usepackage{amsmath}
\begin{document}

\title{Cosmological Constraints on Ghost Dark Energy in the Brans-Dicke Theory by Using MCMC Approach}

\author{Hamzeh Alavirad $^{\ast}$\footnote{hamzeh.alavirad@kit.edu} and Ahmad  Sheykhi$^{\dag, \ddag}$ \footnote{
asheykhi@shirazu.ac.ir}}

\address{$^\ast$ Institute for Theoretical Physics, Karlsruhe Institute of Technology (KIT), 76128 Karlsruhe, Germany\\
          $^\dag$ Physics Department and Biruni Observatory, College of Sciences, Shiraz University, Shiraz 71454, Iran \\
          $^\ddag$ Research Institute for Astronomy and Astrophysics of Maragha (RIAAM), P. O. Box 55134-441, Maragha,
          Iran}
\begin{abstract}
By using a Markov Chain Monte Carlo simulation, we investigate
cosmological constraints on the ghost dark energy (GDE) model in the
framework of the  Brans-Dicke (BD) theory. A combination of the latest
observational data of the cosmic microwave background radiation data from
seven-year WMAP, the baryon acoustic oscillation data form the
SDSS, the supernovae type Ia data  from the Union2 and
the X-ray gas mass fraction data from the Chandra X-ray observations of
the largest relaxed galaxy clusters are used to perform
constraints on GDE in the  BD cosmology. In this paper, we consider both flat and non-flat
universes together with interaction between dark matter  and
dark energy. The main cosmological parameters are obtained
as: $\Omega_{\rm b}h^2= 0.0223^{+0.0016}_{-0.0013}$, $\Omega_{\rm c}h^2=0.1149^{+0.0088}_{-0.0104}$
and $\Omega_{\rm k}=0.0005^{+0.0025}_{-0.0073}$.
In addition, the Brans-Dicke parameter $\omega$ is estimated as $1/\omega\simeq 0.002$. 

\end{abstract}
\maketitle
\section{Introduction}\label{sec:intro}
Accelerating expansion of the Universe \cite{Riess:1998cb, Perlmutter:1998np}
can be explained either by a missing energy component usually
called ``dark energy" (DE) with an exotic equation of state, or by
modifying the underlying theory of gravity on large scales. The
famous examples of the former approach include scalar field models
of DE such as quintessence \cite{Wetterich:1987fm, Ratra:1987rm}, K-essence
\cite{ArmendarizPicon:2000dh, ArmendarizPicon:2000ah}, tachyon \cite{Padmanabhan:2002cp, Sen:2002in}, phantom \cite{Caldwell:1999ew, Nojiri:2003vn, Nojiri:2003jn},
ghost condensate \cite{ArkaniHamed:2003uy, Piazza:2004df}, quintom \cite{Elizalde:2004mq, Nojiri:2005sx, Anisimov:2005ne},
holographic DE \cite{Witten:2000zk}, agegraphic DE \cite{Cai:2007us, Wei:2007ty} and so forth.
For a comprehensive review on DE models, see \cite{Copeland:2006wr, Li:2011sd}.
The latter approach for explanation of the acceleration expansion
is based on the modification of the underlying theory of gravity
on large scales such as $f(R)$ gravity \cite{DeFelice:2010aj} and
braneworld scenarios \cite{Dvali:2000hr, Carena:2006yr, Minamitsuji:2008fz, Sheykhi:2007wn}.

Among various models of DE, the so called ghost dark energy (GDE)
has attracted a lot of interests in recent years. The origin of DE
in this model comes from Veneziano ghosts in QCD theory
\cite{Kawarabayashi:1980dp, Witten:1979vv, Veneziano:1979ec,
Rosenzweig:1979ay}. Indeed, the contribution of the ghosts field
to the vacuum energy in curved space or time-dependent background
can be regarded as a possible candidate for DE
\cite{Ohta:2010in, Urban:2009vy}. The magnitude of this vacuum
energy is of order $\Lambda^3_{QCD} H$, where $H$ is the Hubble
parameter and $\Lambda_{QCD}$ is the QCD mass scale. With
$\Lambda_{\rm QCD}\sim 100 MeV$ and $H \sim 10^{-33}eV$ ,
$\Lambda^3_{\rm QCD}H$ gives the right order of magnitude $\sim
(3\times10^{-3}eV)^4$ for the observed dark energy density
\cite{Ohta:2010in}. The advantages of GDE model compared to other
DE models is that it is totally embedded in standard
model and general relativity, therefore one needs not to introduce any new
parameter, new degree of freedom or to modify gravity. The
dynamical behavior of GDE model in flat universe have been studied
\cite{Cai:2010uf}. The study was also generalized to the universe
with spacial curvature \cite{Sheykhi:2011xz}. The instability of
the GDE model against perturbations was studied in
\cite{Ebrahimi:2011js}. In \cite{Sheykhi:2011fe, Sheykhi:2011nb}
the correspondence between GDE and scalar field models of DE were established. In the presence of bulk viscosity and
varying gravitational constant, the GDE model was
investigated in \cite{Sheykhi:2012zzb}. Other features of the GDE
model have been investigated in Refs. \cite{RozasFernandez:2011je,
Karami:2012yf, KhodamMohammadi:2012ww, Malekjani:2012wc,
Feng:2012gr}. The cosmological constraints on this model have been
considered by some authors \cite{Cai:2010uf, Feng:2012gr,
Cai:2012fq}.

Scalar tensor theories have been reconsidered extensively,
recently. One important example of the scalar tensor theories is
the BD theory of gravity which was introduced by Brans and Dicke
in 1961 to incorporate the Mach's principle in the Einstein's
theory of gravity \cite{Brans:1961sx}. This theory also passes the
observational tests in the solar system domain
\cite{Bertotti:2003rm}. In addition, BD theory can be tested by
the cosmological observations such as the cosmic microwave
background (CMB) and large scale structure (LSS)
\cite{Chen:1999qh, Acquaviva:2007mm, Tsujikawa:2008uc,
Wu:2009zb,Wu:2013}. Since the GDE model have a dynamic behavior,
it is more reasonable to consider this model in a dynamical
framework such as BD theory. It was shown that some features of
GDE in BD cosmology differ from Einstein's gravity
\cite{Ebrahimi:2011ne}. For example, while the original DE is
instable in all range of the parameter spaces in standard
cosmology \cite{Ebrahimi:2011js}, it leads to a stable phase in BD
theory \cite{Saaidi:2012vs}. In the framework of BD cosmology, the
ghost model of DE has been studied \cite{Ebrahimi:2011ne}. It is
also of great interest to see whether the GDE model in the
framework of the BD theory is compatible with observational data
or not.

In this paper, cosmological constraints on  GDE in the BD theory
(GDEBD) \cite{Ebrahimi:2011ne}  theory is performed by using the
Marko Chain Monte Carlo (MCMC) simulation. The used observational
datasets are as follows: cosmic microwave background radiation
(CMB) from WMAP7 \cite{Komatsu:2010fb}, 557 Union2 dataset of type Ia
supernova \cite{Amanullah:2010vv}, baryon acoustic oscillation (BAO) from
SDSS DR7 \cite{Percival:2009xn}, and the cluster X-ray gas mass fraction
from the Chandra X-ray observations \cite{Allen:2007ue}. To put the
constraints, the  modified CosmoMC \cite{cosmomc} code is used.

The organization of this paper is as follows. In section
\ref{sec:GDEBD}, we review the formalism of the GDE in the framework
of Brans-Dicke cosmology. In section \ref{sec:MCMCmethod} the methods
which are used in this paper to analyze the data are introduced.
Section \ref{sec:results} contains the results of the MCMC
simulation and we conclude our paper in section \ref{sec:conc}.

\section{Interacting Ghost Dark Energy in the Brans-Dicke Theory in a Non-Flat Universe}\label{sec:GDEBD}
Let us first review the formalism of the interacting GDE in the
framework of BD theory in a non-flat universe \cite{Ebrahimi:2011ne}. The
action of the  BD theory in the canonical form may be written
\cite{Arik:2005ir}
\begin{equation}
 S=\int{
d^{4}x\sqrt{g}\left(-\frac{1}{8\omega}\phi ^2
{R}+\frac{1}{2}g^{\mu \nu}\partial_{\mu}\phi \partial_{\nu}\phi
+L_M \right)},\label{act1}
\end{equation}
where $R$ is the Ricci scalar and $\phi$ is the BD scalar field.
Varying the action with respect to the metric $g_{\mu\nu}$ and the
BD scalar field $\phi$, yields
\begin{eqnarray}
&&\phi \,G_{\mu\nu}=-8\pi T_{\mu\nu}^{M} - \frac{\omega}{\phi}
\left(\phi_{,\mu}\phi_{,\nu}-\frac{1}{2}g_{\mu\nu}\phi_{,\lambda}\phi^{,\lambda}\right)
-\phi_{;\mu;\nu}+g_{\mu\nu}\Box\phi, \label{eqn1}
\\ & &\Box\phi=\frac{8\pi}{2\omega+3}T_{\lambda}^{M
\,\lambda}, \label{eqn2}
\end{eqnarray}
where $T^M_{\mu\nu}$ stands for the energy-momentum tensor of the
matter fields. The line element of the Friedmann-Robertson-Walker
(FRW) universe is
\begin{eqnarray}
 ds^2=dt^2-a^2(t)\left(\frac{dr^2}{1-kr^2}+r^2d\Omega^2\right),\label{metric}
 \end{eqnarray}
where $a(t)$ is the scale factor, and $k$ is the curvature
parameter with $k = -1, 0, 1$ corresponding to open, flat, and
closed universes, respectively. Nowadays, there are some evidences
in favor of closed universe with a small positive curvature
($\Omega_{\rm k}\simeq0.01$) \cite{Bennett:2012zja}. Using  metric (\ref{metric}),
the field equations (\ref{eqn1}) and (\ref{eqn2}) reduce to
\begin{eqnarray}
&&\frac{3}{4\omega}\phi^2\left(H^2+\frac{k}{a^2}\right)-\frac{1}{2}\dot{\phi}
^2+\frac{3}{2\omega}H
\dot{\phi}\phi=\rho_{\rm m}+\rho_{\rm D},\label{FE1}\\
&&\frac{-1}{4\omega}\phi^2\left(2\frac{{\ddot{a}}}{a}+H^2+\frac{k}{a^2}\right)-\frac{1}{\omega}H
\dot{\phi}\phi -\frac{1}{2\omega}
\ddot{\phi}\phi-\frac{1}{2}\left(1+\frac{1}{\omega}\right)\dot{\phi}^2=p_D,\label{FE2}\\
&&\ddot{\phi}+3H
\dot{\phi}-\frac{3}{2\omega}\left(\frac{{\ddot{a}}}{a}+H^2+\frac{k}{a^2}\right)\phi=0,
\label{FE3}
\end{eqnarray}
where $H=\dot{a}/a$ is the Hubble parameter, $\rho_{\rm D}$ and $p_D$ are,
respectively, the energy density and pressure of DE, and $\rho_{\rm m}$ is
the pressureless matter density which contains both dark matter (DM) and baryonic matter (BM) densities i.e.
$\rho_{\rm m}=\rho_{\rm c}+\rho_{\rm b}$  where $\rho_{\rm c}$ and $\rho_{\rm b}$ are the energy densities of dark matter and baryonic matter
respectively.

To be more general and because of some observational evidences
\cite{Bertolami:2007zm, Tsujikawa:2004dp}, here we propose the case where there is
an interaction between GDE and DM. In this case the
semi-conservation equations read
\begin{eqnarray}
&&
\dot{\rho}_D+3H\rho_{\rm D}(1+w_D)=-Q,\label{consq2}\\
&&\dot{\rho}_{c}+3H\rho_{c}=Q, \label{consm2}\\
&&\dot{\rho}_{b}+3H\rho_{b}=0, \label{consb2}
\end{eqnarray}
where $Q$ represents the interaction term between dark matter and dark energy and here we assume that the baryonic matter is conserved separately. We assume $Q =3\xi
H(\rho_{\rm m}+\rho_{\rm D})$ with $\xi$ being a constant. Such a choice
for interacting term implies the the DE and DM component do not
conserve separately while the total density is still conserved
through
\begin{equation}\label{totcons}
\dot{\rho}+3H(\rho+P)=0,
\end{equation}
where $\rho=\rho_{\rm D}+\rho_{\rm m}$ and $P=P_D$.

\par The ghost energy density
is proportional to the Hubble parameter \cite{Ohta:2010in}
\begin{equation}\label{GDE}
\rho_{\rm D}=\alpha H.
\end{equation}
where $\alpha>0$ is roughly of order $\Lambda_{\rm QCD}^3$ and
$\Lambda_{\rm QCD}$ is QCD mass scale. Taking into account the
fact that $\Lambda_{\rm QCD}\sim 100MeV$ and $H\sim 10^{-33}eV$
for the present time, this gives the right order of magnitude
$\rho_{\rm D}\sim (3\times 10^{-3}\rm {eV})^4$ for the ghost energy
density \cite{Ohta:2010in}.

Since the system of equations (\ref{FE1}-\ref{FE3}) is not closed,
we still have another degree of freedom in analyzing the set of
equations. As usual we assume the BD scalar field $\phi$ has a
power law relation versus the scale factor,
\begin{equation}\label{phipl}
\phi=\phi_0a(t)^{\varepsilon}.
\end{equation}
An interesting case is when $\varepsilon$ is small whereas
$\omega$ is high so that the product $\varepsilon\omega$ results
of order unity \cite{Banerjee:2007zd, Sheykhi:2009dz}. In section \ref{sec:results} we will consider
the $\omega\varepsilon=1$ condition for constraining the model by observational data.
This is interesting because local
astronomical experiments set a very high lower bound on $\omega$
\cite{Will}; in particular, the Cassini experiment implies that
$\omega > 10^4$ \cite{Bertotti:2003rm, Acquaviva:2007mm}. Now we take the time
derivative of relation (\ref{phipl}). We arrive at
\begin{equation}\label{phidot}
 \frac{\dot{\phi}}{\phi}=\varepsilon \frac{\dot{a}}{a}=\varepsilon H.
\end{equation}
Combining Eqs. (\ref{phipl}) and (\ref{phidot}) with the first
Friedmann equation (\ref{FE1}), we get
\begin{equation}\label{frid1}
  H^2(1-\frac{2\omega}{3}\varepsilon^2+2\varepsilon)+\frac{k}{a^2}=\frac{4\omega}{3\phi^2}(\rho_{\rm D}+\rho_{\rm m}).
\end{equation}
As usual the fractional energy densities are defined as
\begin{eqnarray}
\Omega_{\rm c}&=& \frac{\rho_{\rm c}}{\rho_{\mathrm{cr}}}=\frac{4\omega\rho_{\rm c}}{3\phi^2H^2}, \label{Omegac} \\
\Omega_{\rm b}&=& \frac{\rho_{\rm b}}{\rho_{\mathrm{cr}}}=\frac{4\omega\rho_{\rm b}}{3\phi^2H^2}, \label{Omegab} \\
\Omega_{\rm k}&=&\frac{\rho_k}{\rho_{\mathrm{cr}}}=\frac{k}{H^2 a^2},\label{Omegak} \\
\Omega_{\rm D}&=&\frac{\rho_{\rm D}}{\rho_{\mathrm{cr}}}=\frac{4\omega\rho_{\rm D}}{3\phi^2
H^2}, \label{OmegaD}
\end{eqnarray}
where
\begin{eqnarray}\label{rhocr}
\rho_{\mathrm{cr}}=\frac{3\phi^2 H^2}{4\omega}.
\end{eqnarray}
Using (\ref{GDE}) we can rewrite equation (\ref{OmegaD}) as
\begin{equation}\label{OmegaD2}
 \Omega_{\rm D}=\frac{4\omega\alpha}{3\phi^2H}.
\end{equation}
Based on these definitions, equation (\ref{frid1}) can be rewritten as
\begin{equation}\label{frid1omega}
 \gamma=\Omega_{\rm D}+\Omega_{\rm m}-\Omega_{\rm k},
\end{equation}
where $\Omega_{\rm m}=\Omega_{\rm c}+\Omega_{\rm b}$ and we have defined
\begin{equation}
\gamma=1-\frac{2\omega}{3}\varepsilon^2+2\varepsilon.
\end{equation}
Next we take the time derivative of (\ref{frid1}), after using
(\ref{frid1omega}), we find
\begin{equation}\label{doth2}
\frac{\dot{H}}{H^2}=\frac{\Omega_{\rm k}}{\gamma}-(1+\frac{\Omega_{\rm k}}{\gamma})
\left[\varepsilon+\frac{3}{2}+\frac{3}{2}\frac{\Omega_{\rm D}w_D}{\gamma+\Omega_{\rm k}}\right].
\end{equation}
Combining the above equation with Eqs. (\ref{consq2}) and
(\ref{GDE}), we obtain the EoS parameter as
\begin{equation}\label{wDn2}
w_D=-\frac{\gamma}{2\gamma-\Omega_{\rm D}}\left(1-\frac{\Omega_{\rm k}}{3\gamma}-\frac{2\varepsilon}{3}(1+\frac{\Omega_{\rm k}}{\gamma})
+\frac{2\xi}{\Omega_{\rm D}} (\gamma+\Omega_{\rm k}-\Omega_{\rm b})\right).
\end{equation}

The first and second derivatives of the distance can be combined
to obtain the acceleration parameter $q$. It was shown that the
zero redshift value of $q_{0}$, is independent of space curvature,
and can be obtained from the first and second derivatives of the
coordinate distance \cite{Daly:2007dn}. It was argued that $q_{0}$, which
indicates whether the universe is accelerating at the current
epoch, can be obtained directly from the supernova and radio
galaxy data \cite{Daly:2007dn}. The acceleration parameter is given by
\begin{equation}\label{q}
q=-1-\frac{\dot{H}}{H^2}.
\end{equation}
Using (\ref{doth2})the acceleration parameter (\ref{q}) is
obtained as
\begin{equation}\label{q2}
q=(1+\frac{\Omega_{\rm k}}{\gamma})\left[\epsilon+\frac{1}{2}\right]+\frac{\Omega_{D}\epsilon}{2\gamma-\Omega_{\rm D}}
-\frac{3\Omega_{\rm D}}{2(2\gamma-\Omega_{\rm D})}\left[1-\frac{\Omega_{\rm k}}{3\gamma}+\frac{2\xi}{\Omega_{\rm D}}(\gamma+\Omega_{\rm k}-\Omega_{\rm b})\right].
\end{equation}
Finally, we obtain the equation of motions of GDE in BD theory.
For this purpose, we first take the time derivative of relation
(\ref{OmegaD2}). We find
\begin{equation}\label{omegaDdot}
\dot{\Omega}_D=\Omega_{\rm D} H(1+q-2\varepsilon).
\end{equation}
Substituting $q$ from (\ref{q2}) into equation (\ref{omegaDdot}) and
using relation $\Omega_{\rm D}^{\prime}=H\tfrac{d\Omega_{\rm D}}{d\ln{a}}$, we
get
\begin{equation}\label{Omegaprime2n}
\frac{d\Omega_{\rm D}}{d\ln
a}=\Omega_{\rm D}\left[1+(1+\frac{\Omega_{\rm k}}{\gamma})\left[\epsilon+\frac{1}{2}\right]+\frac{\Omega_{D}\epsilon}{2\gamma-\Omega_{\rm D}}
-\frac{3\Omega_{\rm D}}{2(2\gamma-\Omega_{\rm D})}\left[1-\frac{\Omega_{\rm k}}{3\gamma}+\frac{2\xi}{\Omega_{\rm D}}(\gamma+\Omega_{\rm k}-\Omega_{\rm b})\right]-2\varepsilon\right].
\end{equation}
In the remaining part of this paper we will constrain the GDEBD
model by using the most recent observational date in the three different physical models:
model I which is the GDEBD model in a flat universe ($\xi=0$ and
$\Omega_{\rm k}=0$), model II is the  interacting GDEBD model in a flat universe ($\xi\neq0$
and $\Omega_{\rm k}=0$) and finally model III is the  interacting GDEBD model in a non-flat universe
($\xi\neq0$ and $\Omega_{\rm k}\neq0$).

\section{Data fitting method}\label{sec:MCMCmethod}

In this section we discuss the data fitting method in the  Markov Chain Monte Carlo (MCMC) simulation to estimate
the parameters of the  model in section \ref{sec:GDEBD} using cosmological data.

\par To get the best fit values of the relevant parameters,
 the maximum likelihood method is used. The total likelihood
function $\mathcal{L}_{\rm total}=e^{-\chi_{\rm tot}^2/2}$ is defined as the product of
the separate likelihood functions of  uncorrelated observational data with
\begin{equation}
 \chi^2_{\rm tot}=\chi^2_{\rm SNIa}+\chi^2_{\rm CMB}+\chi^{2}_{\rm BAO}+\chi^2_{\rm gas}\;,\label{totchi1}
\end{equation}
where SNIa stands for type Ia supernovae, CMB for cosmic microwave
background radiation, BAO for baryon acoustic oscillation and {\it gas}
stands for X-ray gas mass fraction data. Best fit values
of parameters are obtained by minimizing $\chi_{\rm tot}^2$.
In this paper we use the cosmic microwave background radiation  data
from seven-year WMAP \cite{Komatsu:2010fb},  type Ia supernovae data from 557 Union2
\cite{Amanullah:2010vv}, baryon acoustic oscillation  data from SDSS
DR7 \cite{Percival:2009xn}, and the cluster X-ray gas mass fraction data
from the Chandra X-ray observations \cite{Allen:2007ue}. In the rest of this section we
discuss  each $\chi^2_{i}$ in detail.

To obtain $\chi^2_{\rm CMB}$, we use seven-year WMAP data \cite{Komatsu:2010fb} with the CMB data point
$(R, l_A, z_{\ast})$. The shift
parameter R, which parametrize the changes in the amplitude of the
acoustic peaks is given by \cite{Bond:1997wr}
\begin{equation}
 R=\sqrt{\frac{\Omega_{m0}}{c}}\int_{0}^{z_{\ast}}\frac{dz'}{E(z')}\;,\label{RCMB}
\end{equation}
where $z_{\ast}$ is the redshift of recombination (see (\ref{redshift:recom})), c is the speed of light in vacuum,  $\Omega_{\rm m0}$ is the present
value of the matter density parameter (a ''0`` subscript shows the present value of the related quantity), and $E(z)\equiv H(z)/H_0$. In addition, the acoustic scale $l_{\rm A}$,
which characterizes the changes of the peaks of CMB via the angular diameter distance out to   recombination is
defined as \cite{Bond:1997wr}
\begin{equation}
 l_{\rm A}=\frac{\pi r(z_{\ast})}{r_{\rm s}(z_{\ast})}\;.\label{lCMB}
\end{equation}
The comoving distance $r(z)$ is defined
\begin{equation}
 r(z)=\frac{c}{H_{0}}\int_{0}^{z}\frac{dz'}{E(z')}\;,\label{comdis}
\end{equation}
and  the comoving sound horizon distance at  recombination $r_{\rm s}(z_{\ast})$ is given by
\begin{equation}
 r_{\rm s}(z_{\ast})=\int_{0}^{a(z_{\ast})}\frac{c_{\rm s}(a)}{a^2H(a)}da\;,\label{shdis}
\end{equation}
in terms of  the sound speed $c_{\rm s}(a)$, defined by
\begin{equation}
 c_{\rm s}(a)=\left[3(1+\frac{3\Omega_{\rm b0}}{4\Omega{\gamma0}}a)\right]^{-1/2}\;.\label{soundspeed}
\end{equation}
The seven-year WMAP observations gives $\Omega_{\rm\gamma 0}=2.469\times10^{-5}h^{-2}$
and $\Omega_{\rm b0}= 0.02258^{+0.00057}_{-0.00056}$\cite{Komatsu:2010fb}.
\par The redshift of recombination $z_{\ast}$ is obtained by using the fitting function proposed by Hu and Sugiyama \cite{Hu:1995en}
\begin{equation}
 z_{\ast}=1048[1+0.00124(\Omega_{\rm b0}h^2)^{-0.738}][1+g_{\rm 1}(\Omega_{\rm m0}h^2)^{g_{2}}]\;,\label{redshift:recom}
\end{equation}
where
\begin{equation}
 g_1=\frac{0.0783(\Omega_{\rm b0}h^2)^{-0.238}}{1+39.5(\Omega_{\rm b0}h^2)^{0.763}}, \hspace{1cm} g_2=\frac{0.560}{1+21.1(\Omega_{\rm b0}h^2)^{1.81}}\;.\label{g1g2}
\end{equation}
Then one can define  $\chi^2_{\rm CMB}$  as $\chi^2_{\rm CMB}=X^TC^{-1}_{\rm CMB}X$, with \cite{Komatsu:2010fb}
\begin{subequations}
\begin{align}
 X&=\begin{pmatrix} l_{\rm A}-302.09 \\R-1.725\\z_{\ast}-1091.3 \end{pmatrix},\label{CCMB}\\
 C^{-1}_{\rm CMB}&=\begin{pmatrix} 2.305 & 29.698 &-1.333 \\293689 & 6825.270 & -113.180\\-1.333 & -113.180 & 3.414 \end{pmatrix},\label{invcovCMB}
 \end{align}
\end{subequations}
where $C^{-1}_{\rm CMB}$ is the inverse covariant matrix.

To obtain $\chi^2_{\rm SNIa}$, the SNIa Union2 data \cite{Amanullah:2010vv} is used which includes 577 type Ia supernovae.
The expansion history of the universe $H(z)$ can be given by a specific cosmological model.
To test this model, we can
use the observational data for some predictable cosmological parameter such as luminosity distance $d_{\rm L}$.
Assume that the Hubble parameter $H(z;\alpha_{1},...,\alpha_{n})$ is used to describe the Universe,
where  parameters
$(\alpha_1,...\alpha_n)$ are predicted by a theoretical cosmological model.
  For such a theoretical model we can predict the theoretical
'Hubble-constant free' luminosity distance as
\begin{eqnarray}
 D^{th}_{\rm L}=H_{0}\frac{d_{\rm L}}{c}&=&(1+z)\int_{0}^{z}\frac{dz'}{E(z';\alpha_z,...,\alpha_n)}\nonumber\\&=&H_{\rm0}\frac{1+z}{\sqrt{|\Omega_{\rm k}|}}\mathrm{sinn}\left[\sqrt{|\Omega_{\rm k}|}
\int_{0}^{z}\frac{dz'}{H(z';\alpha_z,...,\alpha_n)}\right],\label{LD}
\end{eqnarray}
where $E\equiv{H}/{H_{0}}$, $z$ is the redshift parameter, and

\[ \mathrm{sinn}(\sqrt{|\Omega_{\rm k}|}x) = \left\{ \begin{array}{ll}
         \sin(\sqrt{|\Omega_{\rm k}|}x) & \mbox{for $\Omega_{\rm k} < 0$}\\
    \sqrt{|\Omega_{\rm k}|}x & \mbox{for $\Omega_{\rm k} = 0$}\\
        \sinh(\sqrt{|\Omega_{\rm k}|}x) & \mbox{for $\Omega_{\rm k} > 0$}.\end{array} \right. \] \label{OmegaK}
Then one can write the theoretical modulus distance
\begin{equation}
 \mu_{\rm th}(z)=5\log_{10}[D_{\rm L}^{\rm th}(z)]+\mu_{0}\; ,\label{muth}
\end{equation}
where $\mu_{0}=5\log_{10}(cH_{0}^{-1}/Mpc)+25$. On the other hand,
the observational modulus distance of SNIa, $\mu_{\rm obs}(z_i)$,
at redshift $z_i$ is given by
\begin{equation}
 \mu_{\rm obs}(z_i)=m_{\rm obs}(z_i)-M,\label{muobs}
\end{equation}
where $m_{\rm obs}$ and $M$ are apparent and absolute magnitudes of SNIa respectively.
Then the parameters of the theoretical model, $\alpha_{i}$s, can be determined
by a likelihood analysis by defining $\bar{\chi}^2_{\rm SNIa}(\alpha_i, M')$ in  (\ref{totchi1}) as
\begin{eqnarray}
 \bar{\chi}^2_{\rm SNIa}(\alpha_i, M')&\equiv&\sum_{\rm j}\frac{(\mu_{\rm obs}(z_j)-\mu_{\rm th}(\alpha_{\rm i},z_j))^2}{\sigma_{\rm j}^2}\\ \nonumber
&=&\sum_{\rm j}\frac{(5\log_{10}[D_{\rm L}(\alpha_{\rm i},z_j)]-m_{\rm obs}(z_j)+M')^2}{\sigma_{\rm j}^2}\;
,
\end{eqnarray}
where the nuisance parameter, $M'=\mu_{0}+M$, can be marginalized over as
\begin{equation}
 {\chi}^{2}_{\rm SNIa}(\alpha_{\rm i})=-2\ln\int_{\rm -\infty}^{+\infty}dM'\exp[-\frac{1}{2}\chi^2_{\rm SNIa}(\alpha_i, M')]\;\label{chiSNf} .
\end{equation}

Here we should mention an important point about using supernovae data
 to constrain the Brans-Dicke theories which have a varying gravitational coupling constant.
Variations of gravitational coupling constant and apparent magnitude of
supernovae are correlated as follows.  The  luminosity $L$  of a supernova is powered by Nickel-35 mass which is
proportional to the Chandrasekhar mass
\begin{equation}
L_{\rm SN}\sim M_{\rm CH}\sim G^{-3/2}.
\end{equation}
Moreover, the luminosity distance $d_{L}$
 is the integral over the inverse Hubble parameter, which
is proportional to $G^{-1/2}$.
Therefore, the apparent  magnitude $m_{\rm obs}$
\begin{equation}
m_{\rm obs}=-2.5{\rm Log} L + 5{\rm Log} d_{L}
\end{equation}
varies with a change in the gravitational coupling constant $\Delta G$ as
\begin{equation}
 \Delta m_{\rm obs} \sim -\frac{1}{8}\frac{\Delta G}{G}.\label{varG1}
\end{equation}
On the other hand, in the Brans-Dicke theory we have
\begin{equation}
 \frac{\dot{G}}{G}=-\frac{\phi}{\dot{\phi}}
\end{equation}
where $\phi$ is the Brans-Dicke scalar field. In the slow roll approximation
we can write
\begin{equation}
 \dot{\phi}=H\epsilon(1-q)
\end{equation}
where $q$ is the decelaration parameter.
The average value of $q$ between today and $z\sim1.2$ (the redshift when the SN measurement are probing
the dark energy) is of order unity, and by using $H=d\ln a/dt$, we can write
\begin{equation}
 \frac{\Delta G}{G}\sim -\epsilon\Delta\ln a.\label{varG2}
\end{equation}
By using Eqs. (\ref{varG1}) and (\ref{varG2}) we obtain
\begin{equation}
 \Delta m_{\rm obs} \sim\frac{\epsilon}{8}\Delta\ln a.
\end{equation}
In Union2 data set, the redshift interval is between $0.51$ and $1.12$, i.e. $\Delta\ln a\sim1$,
with the systematic error of order $0.03$
in the measurement of apparent magnitude. Therefore,  the systematic error can induce
a bias roughly of order $0.3$ on parameter $\epsilon$, which is three order of magnitudes larger than the statistical errors,
as we will discuss in the  next section.
Therefore, in order to constrain $\epsilon$ with a higher precision, we combine the supernovae data
with other cosmological data sets as follows.  For more detailed discussion on possible evolution of
the gravitational constant from cosmological type Ia supernovae see \cite{Gaztanaga:2001fh}.

The baryon acoustic oscillation data from the Sloan Digital Sky Survey
(SDSS) Data Release 7 (DR7) \cite{Percival:2009xn} is used here for
constraining model parameters. The data constrain parameter
$d_{z}\equiv r_{\rm s}(z_{\rm d})/D_{\rm V}(z)$, where $r_{\rm s}(z_{\rm d})$ is the comoving sound
 horizon distance (see (\ref{shdis})) at the drag epoch  (where baryons were released from photons) and $D_{\rm V}$ is given by \cite{Eisenstein:2005su}
\begin{equation}
D_{\rm V}(z)\equiv\left[c\left(\int_{0}^{z}\frac{dz'}{H(z')}\right)^2\frac{z}{H(z)}\right]^{1/3}\;.\label{DBAO}
\end{equation}
The drag redshift is given by the fitting formula \cite{Eisenstein:1997ik}
\begin{equation}
 z_{\rm d}=\frac{1291(\Omega_{\rm m0}h^2)^{0.251}}{1+0.659(\Omega_{\rm m0}h^2)^{0.828}}\left[1+b_{1}(\Omega_{\rm b0}h^2)^{b_{2}}\right]\; ,\label{dragredshift}
\end{equation}
where
\begin{eqnarray}
 b_{1}&=&0.313(\Omega_{\rm m0}h^2)^{-0.419}[1+0.607(\Omega_{\rm m0}h^2)^{0.607}]\;,\nonumber\\ b_{2}&=&0.238(\Omega_{\rm m0}h^2)^{0.223}\; . \label{b1b2bao}
\end{eqnarray}
Then we can obtain $\chi^2_{\rm BAO}$ by $\chi^2_{\rm BAO}=Y^TC^{-1}_{\rm BAO}Y$, where
\begin{equation}
 Y=
\begin{pmatrix}
 d_{0.2}-0.1905\\d_{0.35}-0.1097\end{pmatrix}\; ,\label{YBAO}
\end{equation}
 and its covariance matrix is given by \cite{Percival:2009xn}
\begin{equation}
 C^{-1}_{\rm BAO}=\begin{pmatrix}30124 & -17227\\-17227 & 86977\end{pmatrix}\; .\label{CBAO}
\end{equation}

The ratio of X-ray gas mass to the total mass of a cluster is defined
as the X-ray gas mass fraction  \cite{Allen:2007ue}.
  The model fitted to the $\Lambda$CDM model is \cite{Allen:2007ue}
\begin{eqnarray}
f_{\rm gas}^{\Lambda CDM}(z)=\frac{K A \gamma b(z)}{1+s(z)}\left(\frac{\Omega_b}{\Omega_{\rm m0}}\right)
\left(\frac{D_A^{\Lambda CDM}(z)}{D_A(z)}\right)^{1.5}\; .
\label{fgasLCDM}
\end{eqnarray}
The elements in equation (\ref{fgasLCDM}) are defined as follows:
 $D_{\rm A}^{\Lambda CDM} (z)$ and $D_{\rm A}(z)$ are
the proper angular diameter distance in  $\Lambda$CDM
and the alternative theoretical model respectively, where
\begin{eqnarray}
D_A(z)=\frac{c}{(1+z)\sqrt{|\Omega_{\rm k}|}}\mathrm{sinn}\left[\sqrt{|\Omega_{\rm k}|}\int_0^z\frac{dz'}{H(z')}\right]\;
.
\end{eqnarray}
The angular correction
factor $A$
\begin{eqnarray}
A=\left(\frac{\theta_{2500}^{\Lambda CDM}}{\theta_{2500}}\right)^\eta \approx
\left(\frac{H(z)D_A(z)}{[H(z)D_A(z)]^{\Lambda CDM}}\right)^\eta\; ,
\end{eqnarray}
is caused by changes in angle for the alternative theoretical
model $\theta_{2500}$ compared to $\theta_{2500}^{\Lambda CDM}$,
where $\eta=0.214\pm0.022$ \cite{Allen:2007ue} is the slope of the $f_{\rm gas}(r/r_{2500})$
data within the radius $r_{2500}$
($r_{2500}$ is the radius of the gas core in ${\rm Mpc/h}$ units).

The bias factor $b(z)$ in equation  (\ref{fgasLCDM}) contains
information about the uncertainties in the cluster depletion
factor $b(z)= b_0(1+\alpha_b z)$ and  the parameter $\gamma$ accounts
for departures from the hydrostatic equilibrium. The function
$s(z)=s_0(1 +\alpha_s z)$ denotes the uncertainties of the
baryonic mass fraction in stars with a Gaussian prior for $s_0$,
with $s_0=(0.16\pm0.05)h_{70}^{0.5}$ \cite{Allen:2007ue}. The factor
$K$ describes the combined effects of the residual uncertainties,
such as the instrumental calibration. A Gaussian prior for the
'calibration' factor is considered as $K=1.0\pm0.1$ \cite{Allen:2007ue}.
\par Then $\chi^2_{\rm gas}$ is defined as \cite{Allen:2007ue}
\begin{eqnarray}
\chi^2_{\rm gas}&=&\sum_i^N\frac{[f_{\rm gas}^{\Lambda
CDM}(z_i)-f_{\rm gas}(z_i)]^2}{\sigma_{\rm f_{\rm gas}}^2(z_i)}+\frac{(s_{0}-0.16)^{2}}{0.0016^{2}}
\\ \nonumber &&+\frac{(K-1.0)^{2}}{0.01^{2}}+\frac{(\eta-0.214)^{2}}{0.022^{2}}\;,\label{chifgas}
\end{eqnarray}
with the statistical uncertainties $\sigma_{\rm f_{\rm gas}}(z_{\rm i})$ and
\begin{eqnarray}
f_{\rm gas}(z)=\frac{K A \gamma b(z)}{1+s(z)}\left(\frac{\Omega_b}{\Omega_{\rm m0}}\right)
\left(\frac{D_A^{\Lambda CDM}(z)}{D_A(z)}\right)^{1.5}\; .
\label{fgasLCDM}
\end{eqnarray}

\par At the end of this section we should assert that the data points parameters of the CMB and BAO data sets which we use in this paper
 are the best fit values for  $\Lambda$CDM and the error estimates are also based on the $\Lambda$CDM model.
 Therefore they are not completely accurate in this application. However they are the only parameters
 which we have to constrain our model.

\section{Results}\label{sec:results}
Finally we apply a Markov chain Monte Carlo simulation on the parameters of the
GDEBD model by using the publicly available CosmoMC code \cite{Lewis:2002ah}.
The parameter vectors are
$P^{\rm I}_{\rm s}=\{\Omega_{\rm b}h^2,\Omega_{\rm c}h^2,\epsilon\}$,
$P^{\rm II}_{\rm s}=\{\Omega_{\rm b}h^2,\Omega_{\rm c}h^2,\epsilon, \xi\}$,
$P^{\rm III}_{\rm s}=\{\Omega_{\rm b}h^2,\Omega_{\rm c}h^2,\epsilon,\xi, \Omega_{\rm k}\}$ for
the flat non-interacting (model I), flat interacting (model II) and non-flat interacting (model III) models respectively.
The basic cosmological parameters are taken in the following priors:
$\Omega_{\rm b}h^2=[0.005 0.9]$, $\Omega_{\rm c}h^2=[0.01,\; 0.99]$ and $\Omega_{\rm k}=[-0.05,\; 0.05]$.
In addition, as we mentioned in section \ref{sec:GDEBD}, for the model fitting, we consider $\omega\varepsilon=1$ condition.
The results are presented in table \ref{tab:MCMC}.

\par From table \ref{tab:MCMC} one can see that the main cosmological parameters $\Omega_{\rm b}h^2$, $\Omega_{\rm c}h^2 $, $\Omega_{\rm DE}$ and  $\Omega_{\rm k}$
in all three models are compatible with the results of the
$\Lambda$CDM model \cite{Bennett:2012zja}. In addition, in the
presence of interaction between DM and DE, the parameter
$\epsilon$ decreases and so the Brans-Dicke parameter $\omega$
increases. The best fit value of the parameter $\epsilon=1/\omega$
in all three models is compatible with the results of other
cosmological constraining works  (however see the discussion
following Eq. (\ref{chiSNf})). For example in  \cite{Chen:1999qh}
the authors by using the CMB temperature and polarization
anisotropy data, found $1/\omega\simeq 0.001$. Wu and Chen in
\cite{Wu:2009zb,Wu:2013} by using the five-year WMAP and SDSS data
obtained $\omega>97.8$. For other cosmological constraints on the
BD theory see \cite{ Tsujikawa:2008uc, Nagata:2002tm,
Nagata:2003qn, Acquaviva:2004ti, Schimd:2004nq, Acquaviva:2007mm}.
This estimated value is also compatible with the results of the
solar system tests of the scalar-tensor theories such as the
Cassini experiment where it has been obtained $\omega > 10^4$
\cite{Bertotti:2003rm, Acquaviva:2007mm}.
The positive best fit value of parameter $\xi$ describe a conversion of dark matter
to dark energy although both in flat and non-flat universes, in 1-$\sigma$ CL, an
inverse conversion is possible as well. The interacting DE and DM models have been
constrained by observational data by many authors with different parametrization of the
interacting parameter $Q$ \cite{Chimento:2003iea, Zimdahl:2002zb, Valiviita:2009nu,
Pettorino:2012ts, Salvatelli:2013wra, He:2010im, Clemson:2011an}. He et al. in \cite{He:2010im}
have parametrized the interaction parameter as in the present paper although they have chosen
the prior on parameter $\xi$ as $\xi=[0, 0.02]$. They obtained the best fit value
of parameter $\xi$ as $\xi=0.0006\pm0.0006$.

\begin{table*}[h!]
\begin{center}
    \begin{tabular}{|c||c|c|c|}
        \hline
        Parameter     & Non-interacting & Interacting & Interacting non-flat\\ \hline\hline
        $\Omega_{\rm b}h^2$ & ~   $0.0223^{+0.0016+0.0019}_{-0.0013-0.0018}$           & ~  $0.0224^{+0.0016+0.0018}_{-0.0016-0.0020}$ & ~   $0.0223^{+0.0018+0.0020}_{-0.0013-0.0017}$\\ \hline
        $\Omega_{\rm c}h^2 $& ~    $0.1149^{+0.0088+0.0104}_{-0.0104-0.0119}$          & ~   $0.1118^{+0.0117+0.0139}_{-0.0087-0.0101}$ & ~ $0.1151^{+0.0089+0.0098}_{-0.0132-0.0151}$  \\ \hline
        $\Omega_{\rm DE}$  & ~      $0.7148^{+0.0445+0.0535}_{-0.0356-0.0464}$      & ~ $0.7291^{+0.0360+0.0451}_{-0.0550-0.0700}$ & ~  $0.7133^{+0.0472+0.0552}_{-0.0386-0.0430}$ \\ \hline
        $\Omega_{\rm k}$ & ~ \ldots &~ \ldots &~ $0.0005^{+0.0025+0.0026}_{-0.0073-0.0073}$\\ \hline
    $\epsilon$         & ~       $0.0020^{+0.0004}_{-0.0006}$     & ~   $0.0017^{+0.0008}_{-0.0003}$ & ~ $0.0017^{+0.0007}_{-0.0003}$ \\ \hline
    $\xi$           &~ \ldots                           & ~ $0.4895^{+0.3662+0.3769}_{-0.5951-0.5951}$ & ~  $0.6004^{+0.3638+0.3638}_{-0.6031-0.6065}$\\ \hline
        $H_{0}$         & ~      $69.3902^{+4.3469+5.0061}_{-3.1604-3.6291}$        & ~ $70.4046^{+3.5815+4.1015}_{-4.2224-5.0097}$  & ~  $69.3060^{+4.2517+4.6106}_{-3.7536-4.5922}$\\ \hline
        \end{tabular}
\caption{The best fit values of the cosmological and model parameters in the  the GDE model in the  BD theory with $1\sigma$ and $2\sigma$ regions. Here CMB, SNIa and BAO and
X-ray mass gas fraction data together with the BBN constraints have been used.}
\end{center}\label{tab:MCMC}
\end{table*}

\begin{figure*}[t!]
\centering
\begin{tabular}{ccc}
\epsfig{file=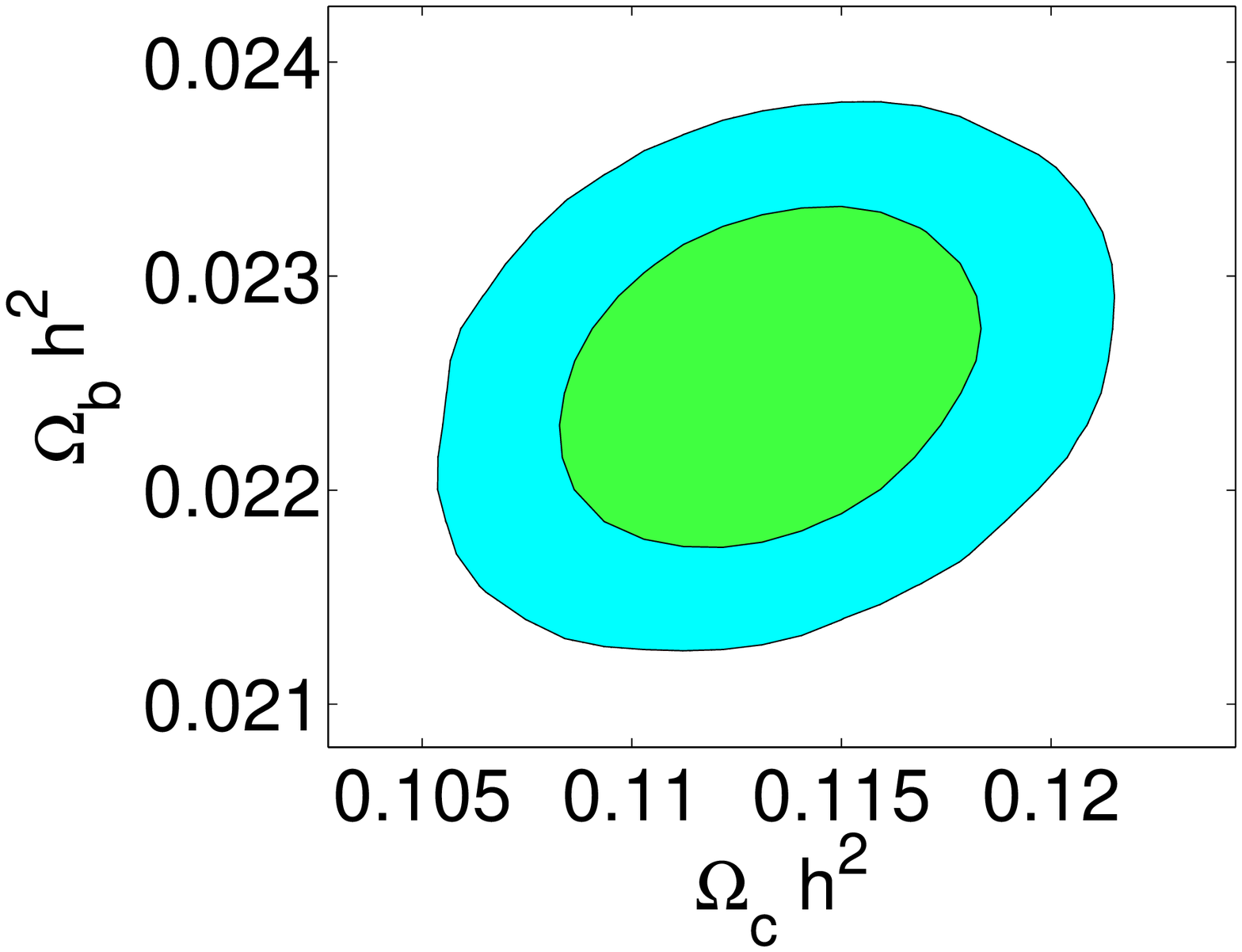,width=0.18\linewidth} & &  \\
\epsfig{file=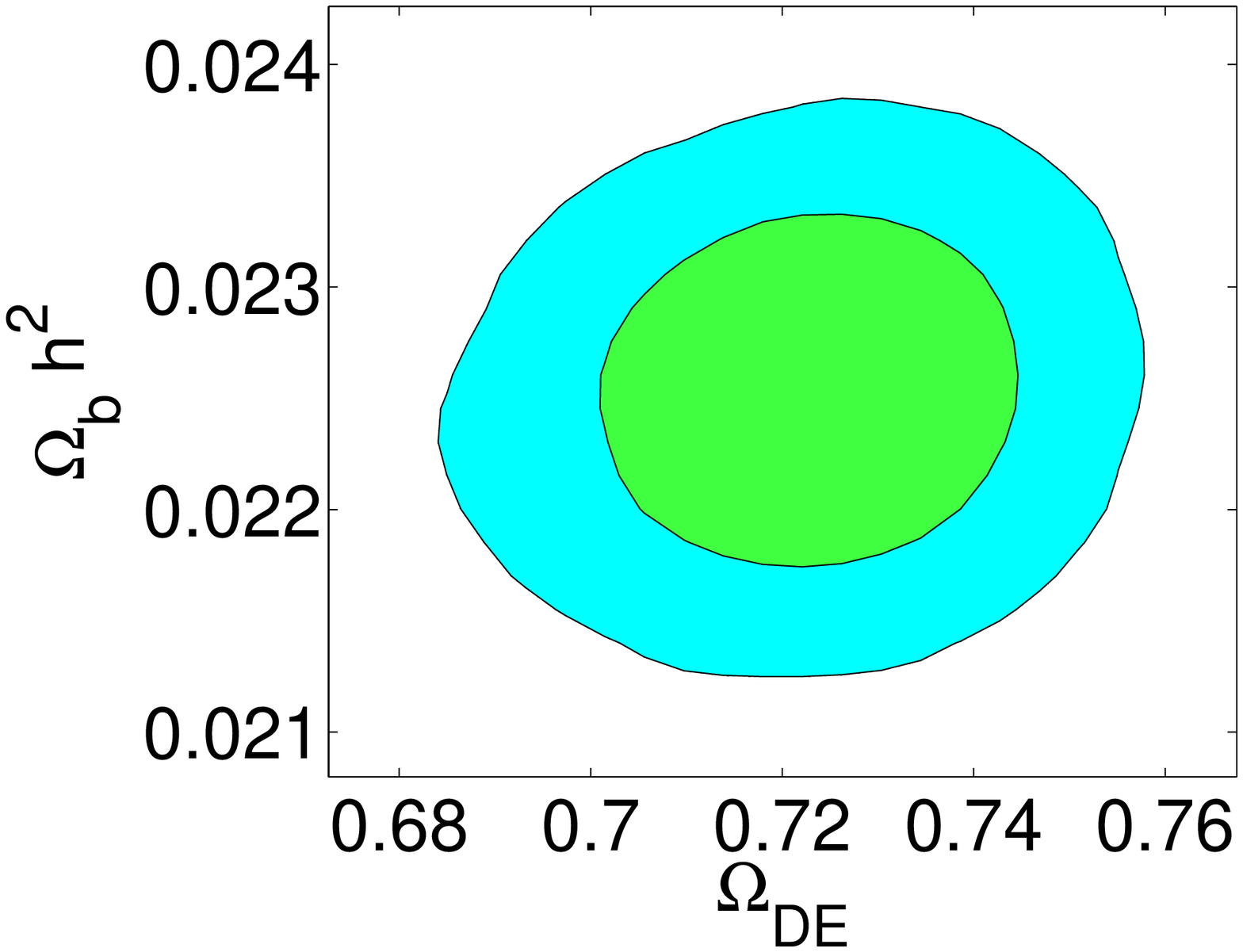,width=0.18\linewidth} & \epsfig{file=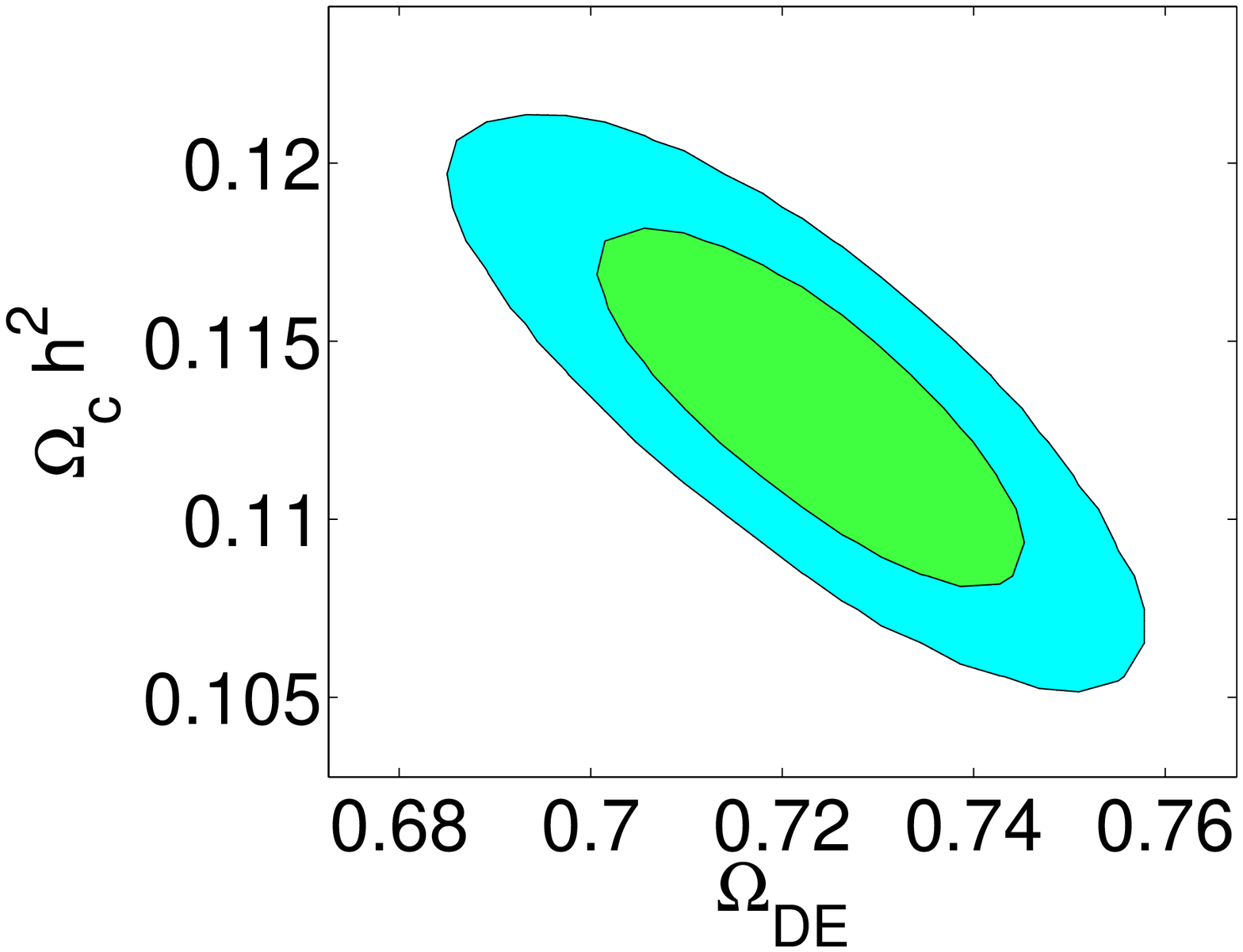,width=0.18\linewidth}& \\
\epsfig{file=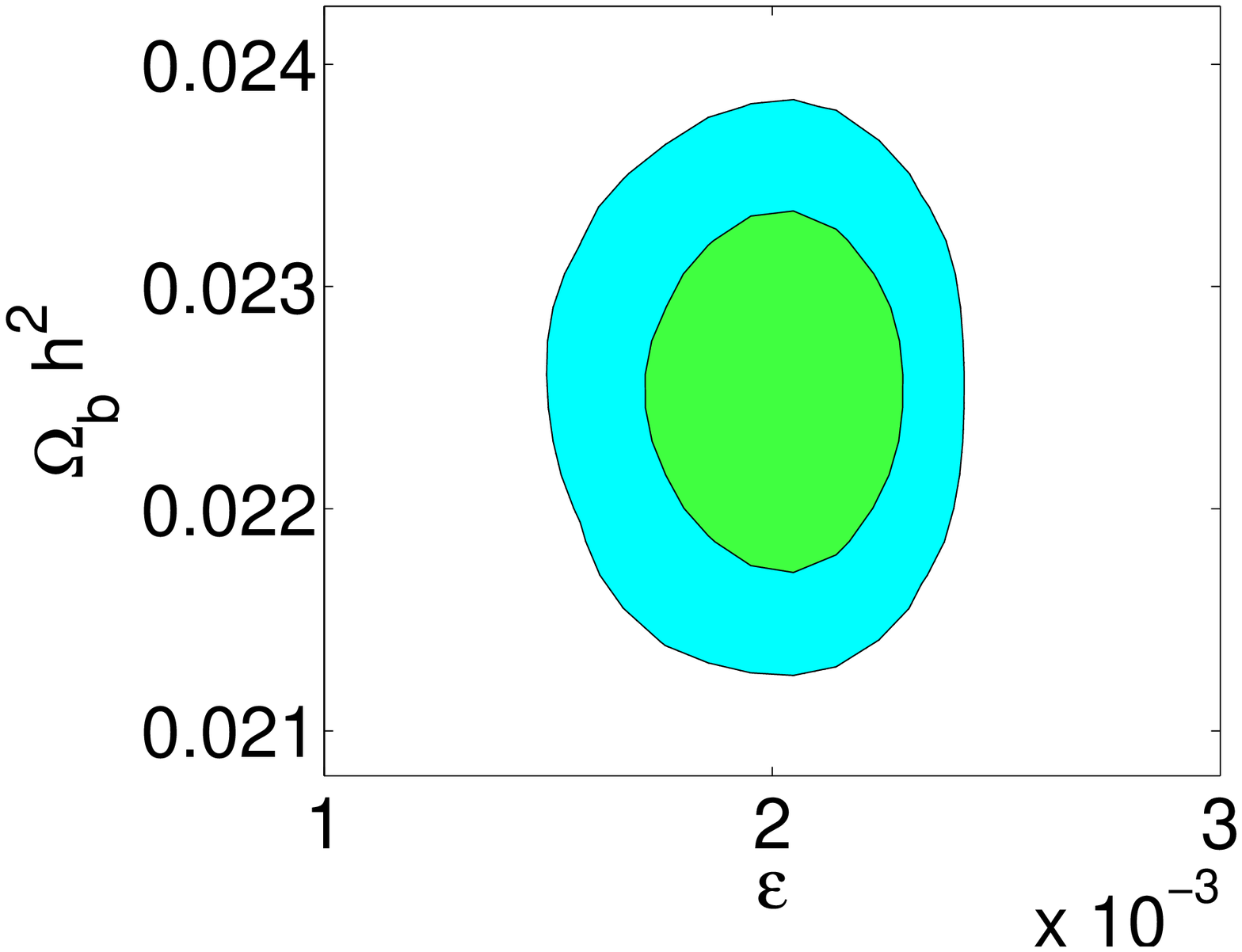,width=0.18\linewidth} &\epsfig{file=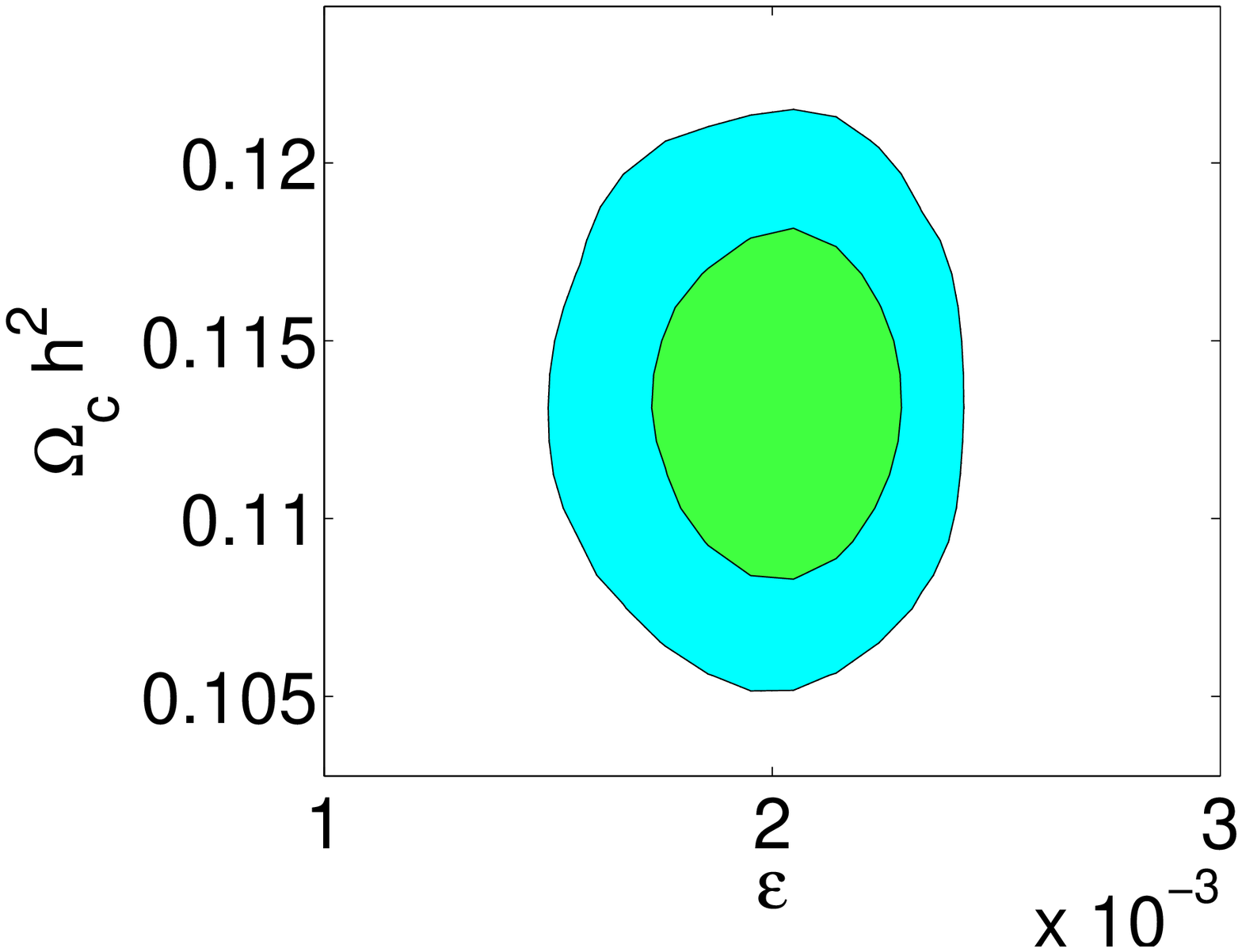,width=0.18\linewidth} & \epsfig{file=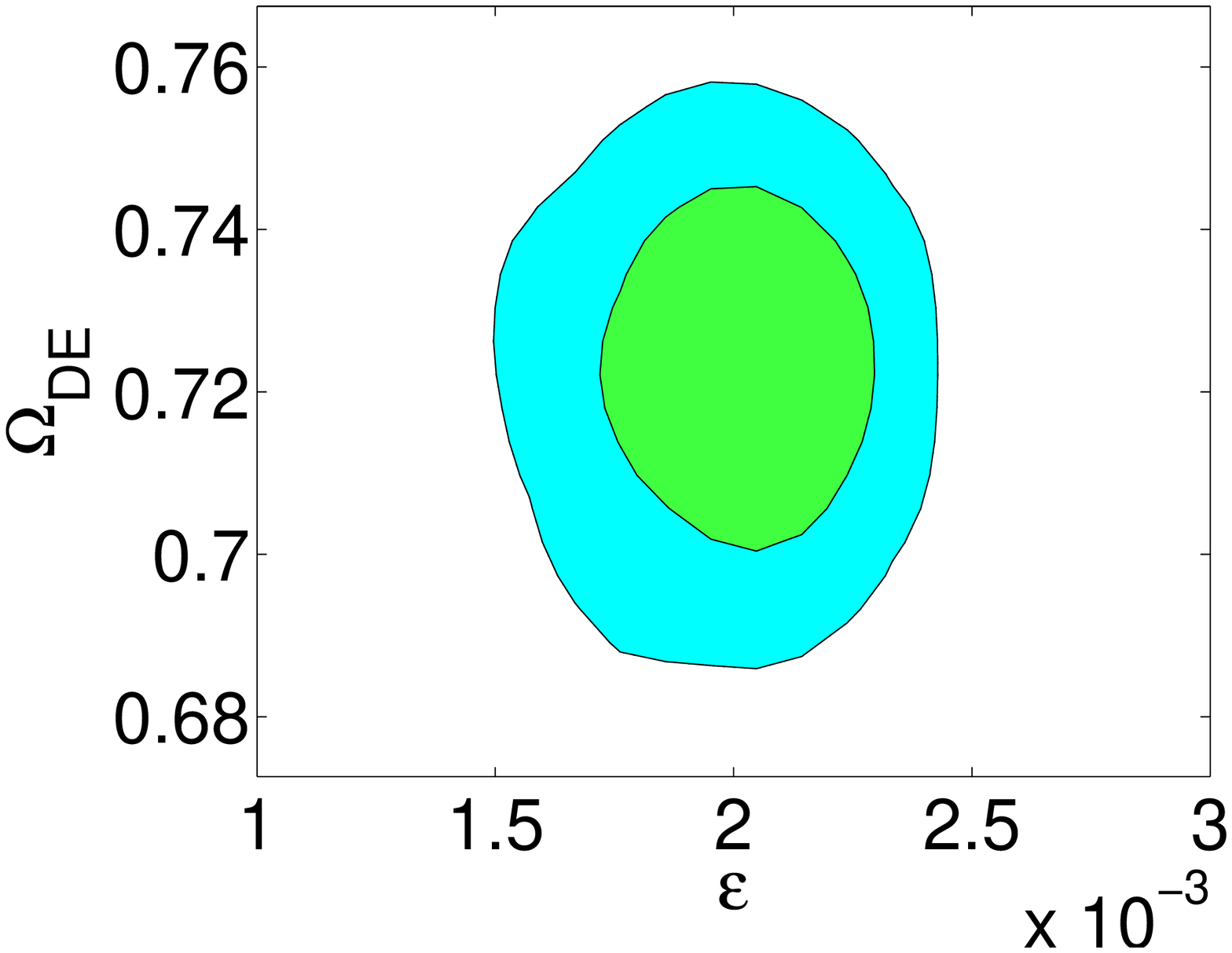,width=0.18\linewidth}\\
\end{tabular}
\caption
{2-dimensional constraint of the cosmological and model parameters contours
in the flat non-interacting GDE model in the  BD theory with $1\sigma$ and $2\sigma$ regions. To produce these plots,
SNIa+CMB+BAO+X-ray gas mass fraction data together with the BBN constraints have been used.}\label{fig:FNI}
\end{figure*}

%


\begin{figure*}[t]
\centering
\begin{tabular}{cccc}
\epsfig{file=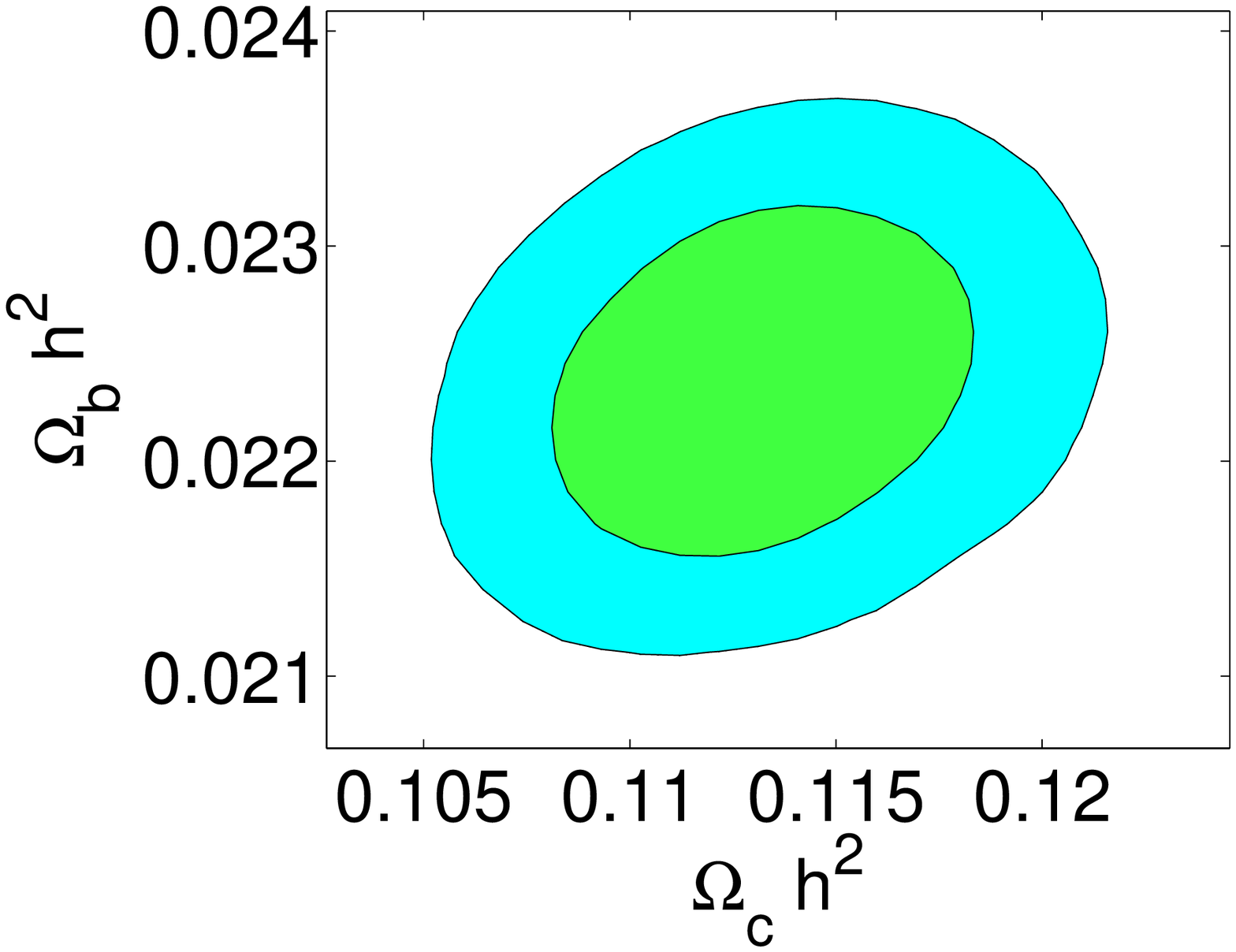,width=0.18\linewidth} & & &  \\
\epsfig{file=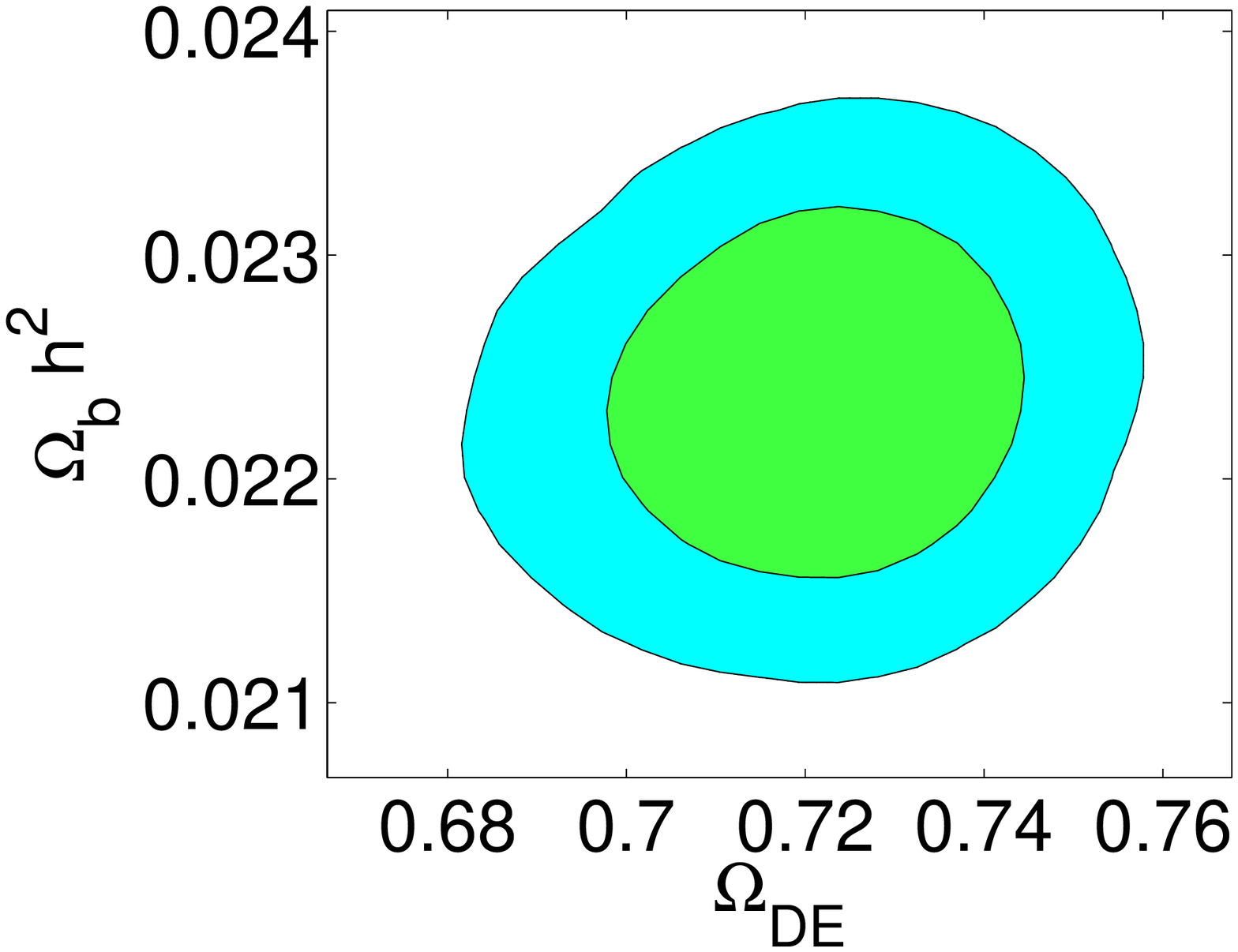,width=0.18\linewidth} & \epsfig{file=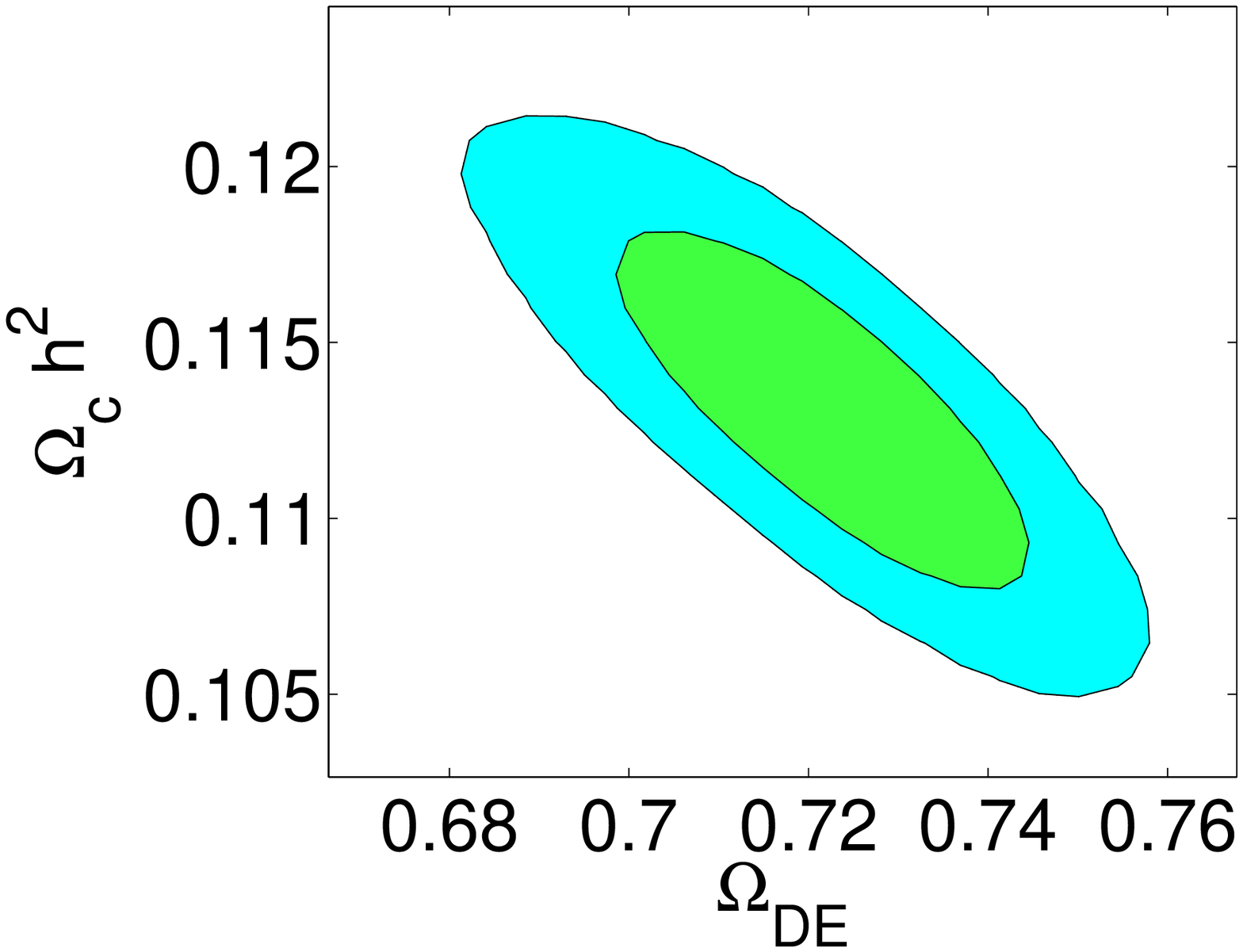,width=0.18\linewidth}& & \\
\epsfig{file=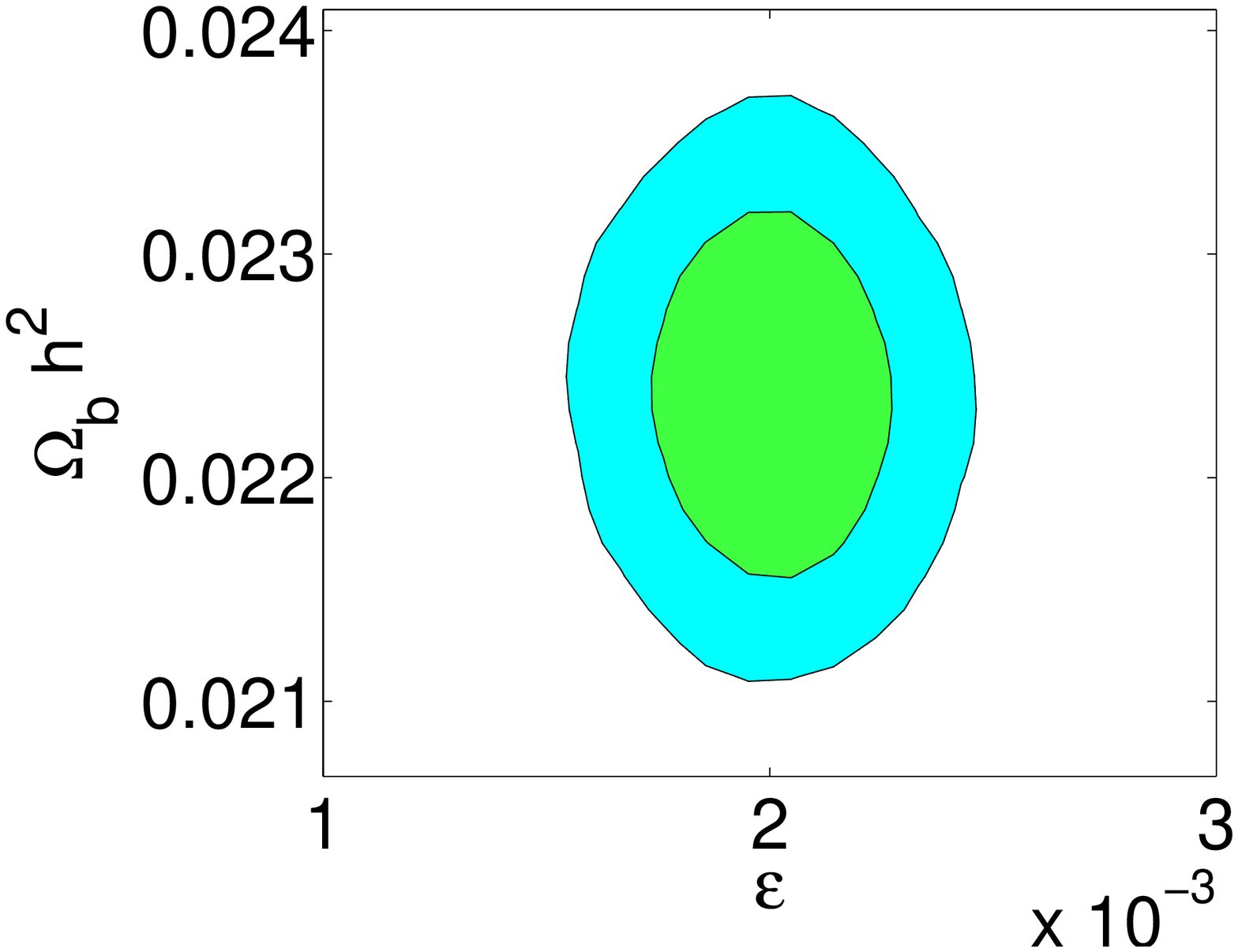,width=0.18\linewidth} &\epsfig{file=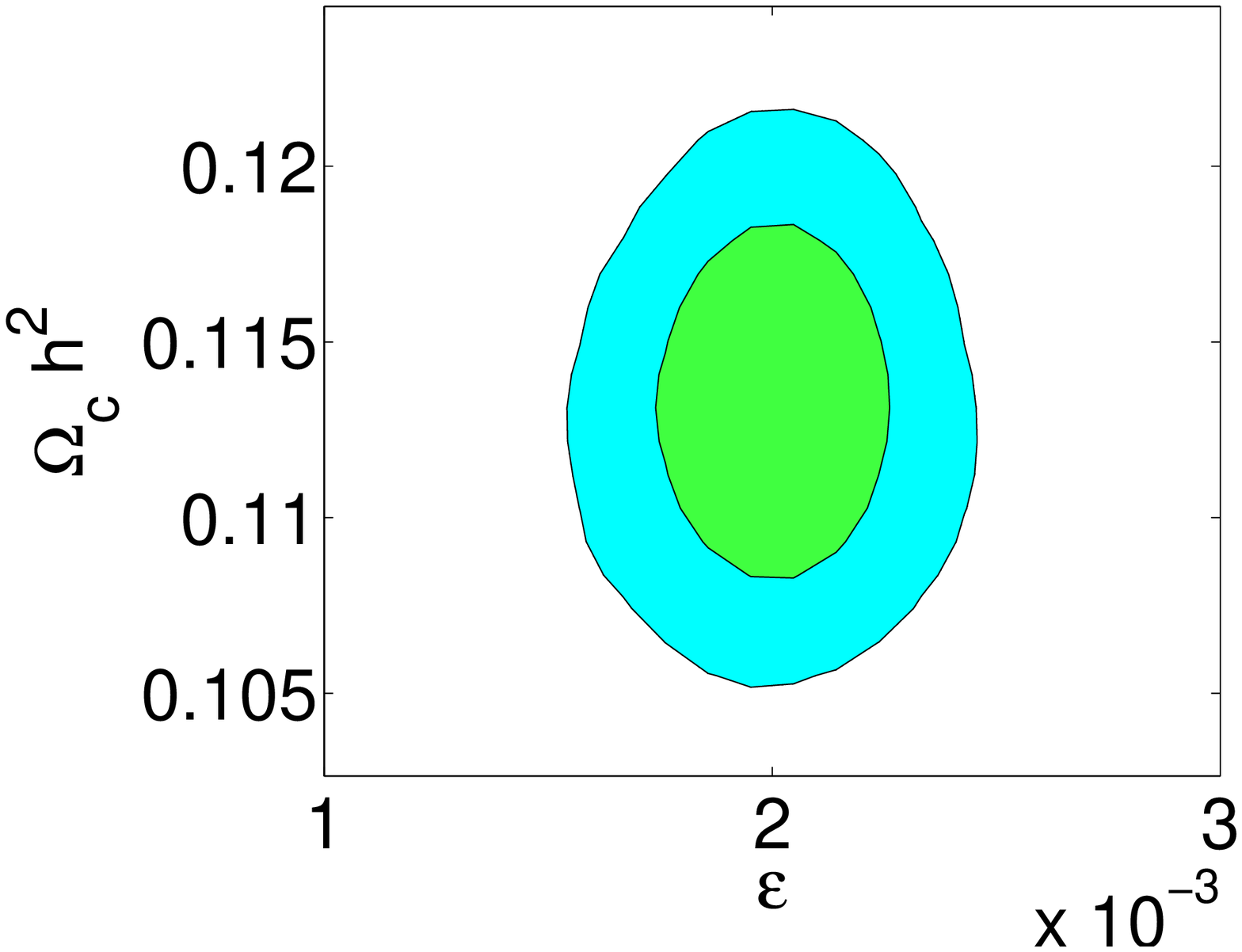,width=0.18\linewidth} & \epsfig{file=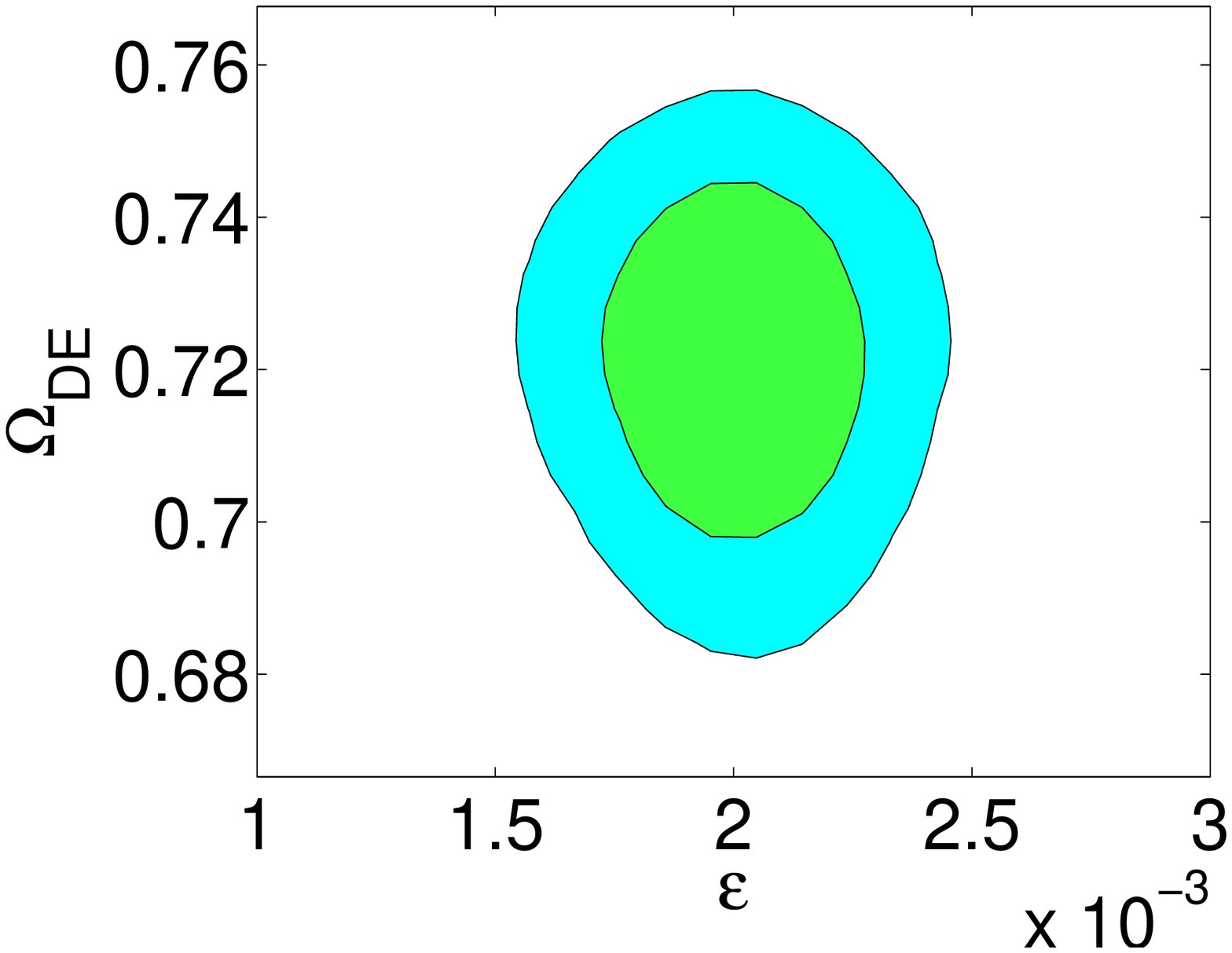,width=0.18\linewidth}& \\
\epsfig{file=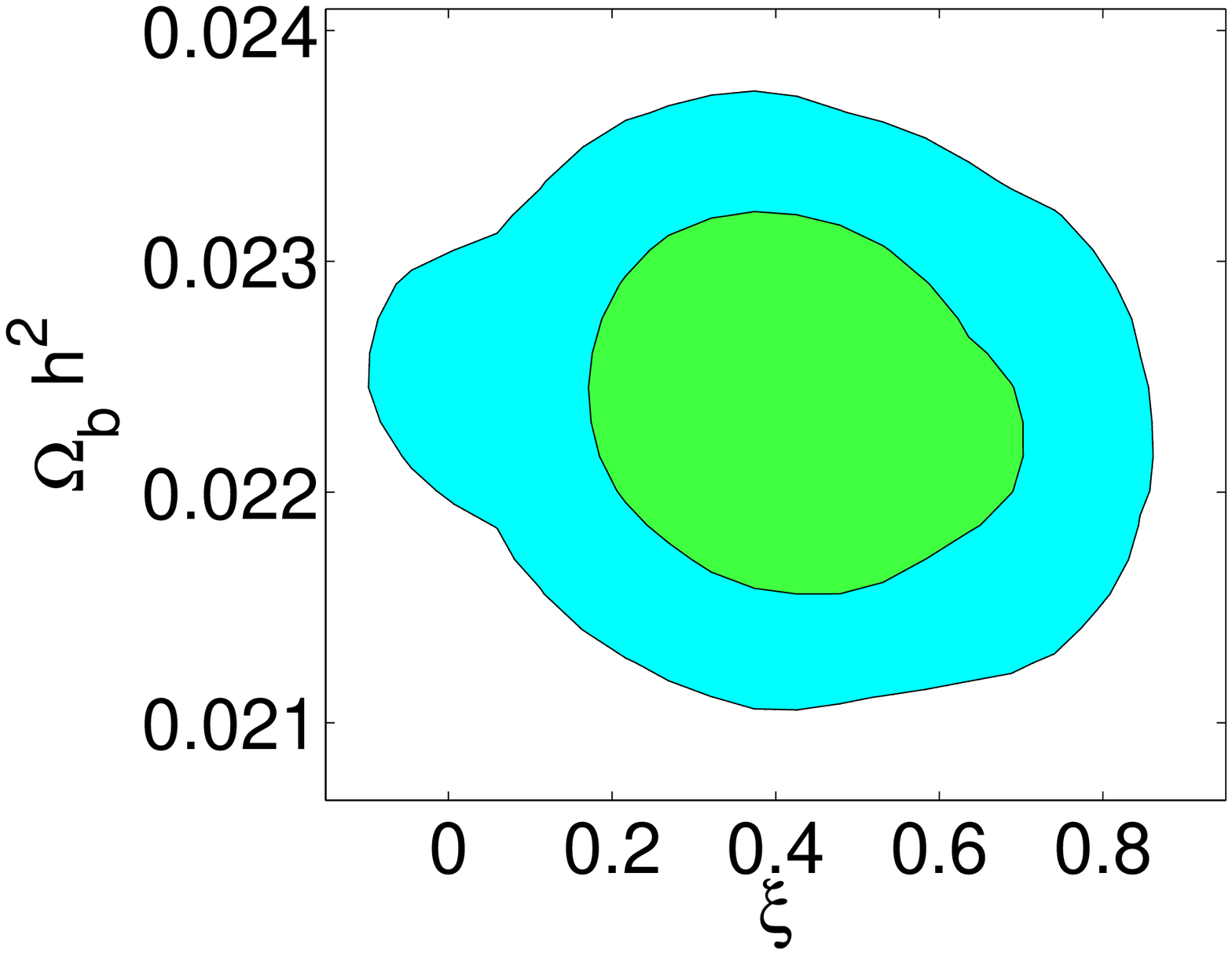,width=0.18\linewidth} &\epsfig{file=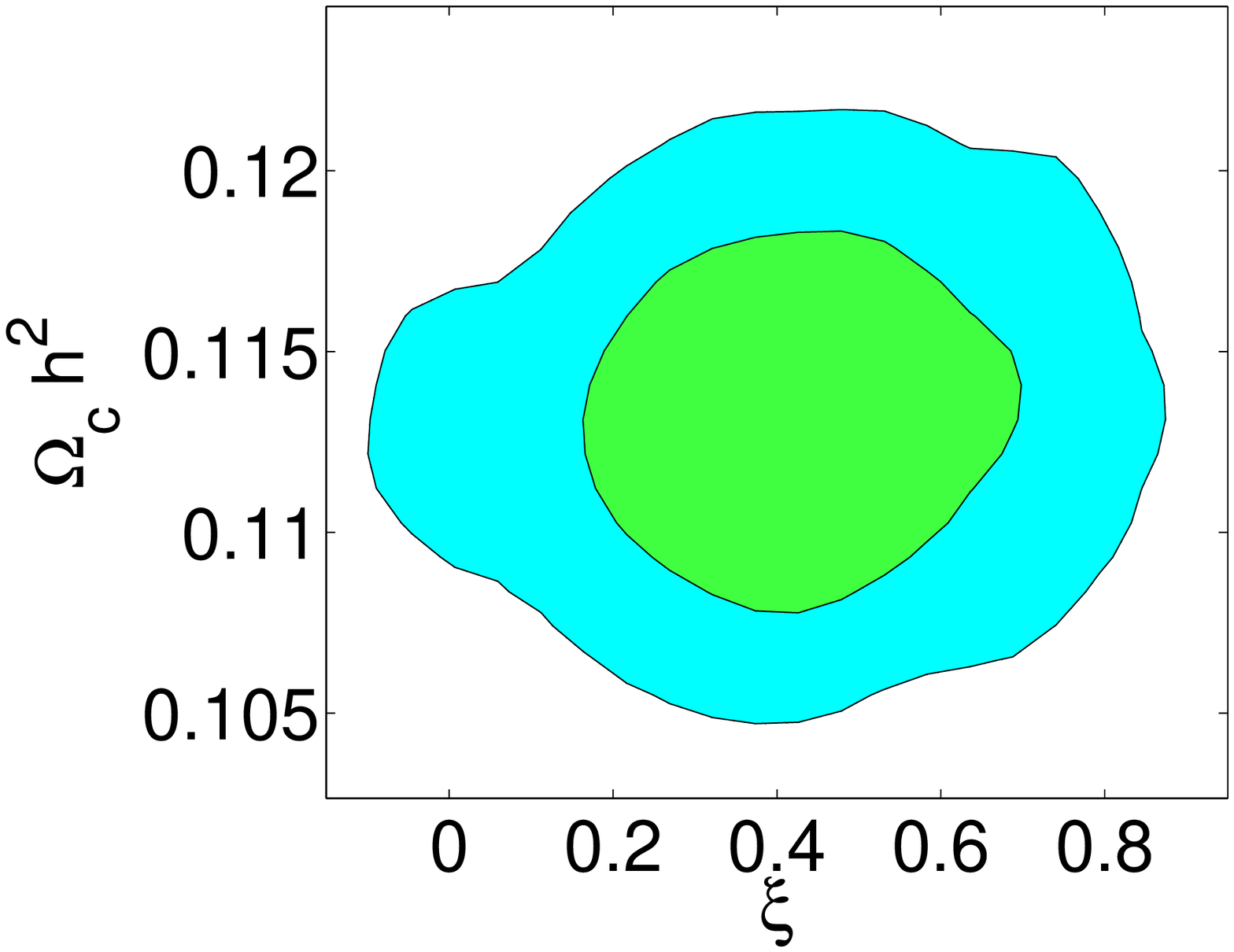,width=0.18\linewidth} & \epsfig{file=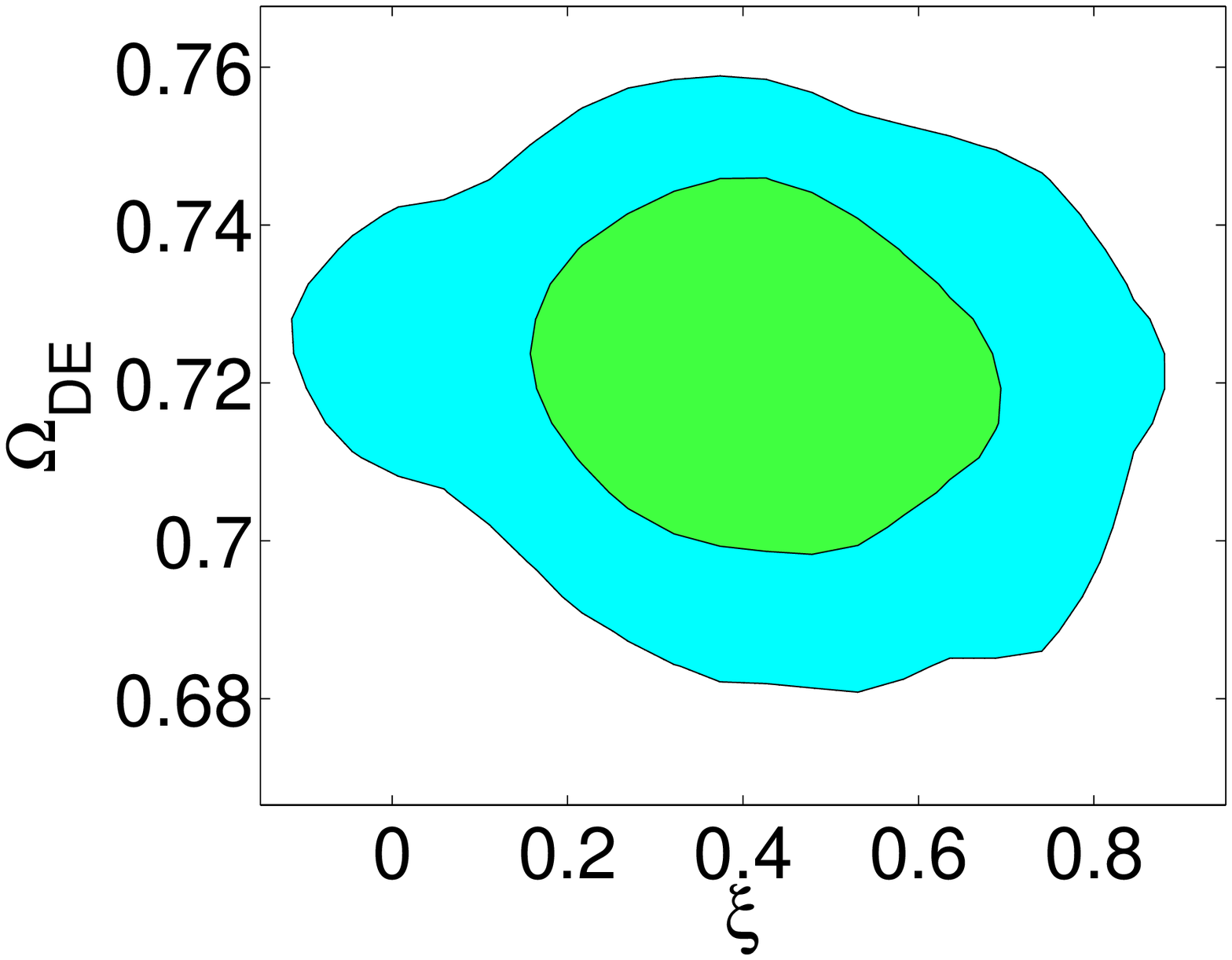,width=0.18\linewidth}& \epsfig{file=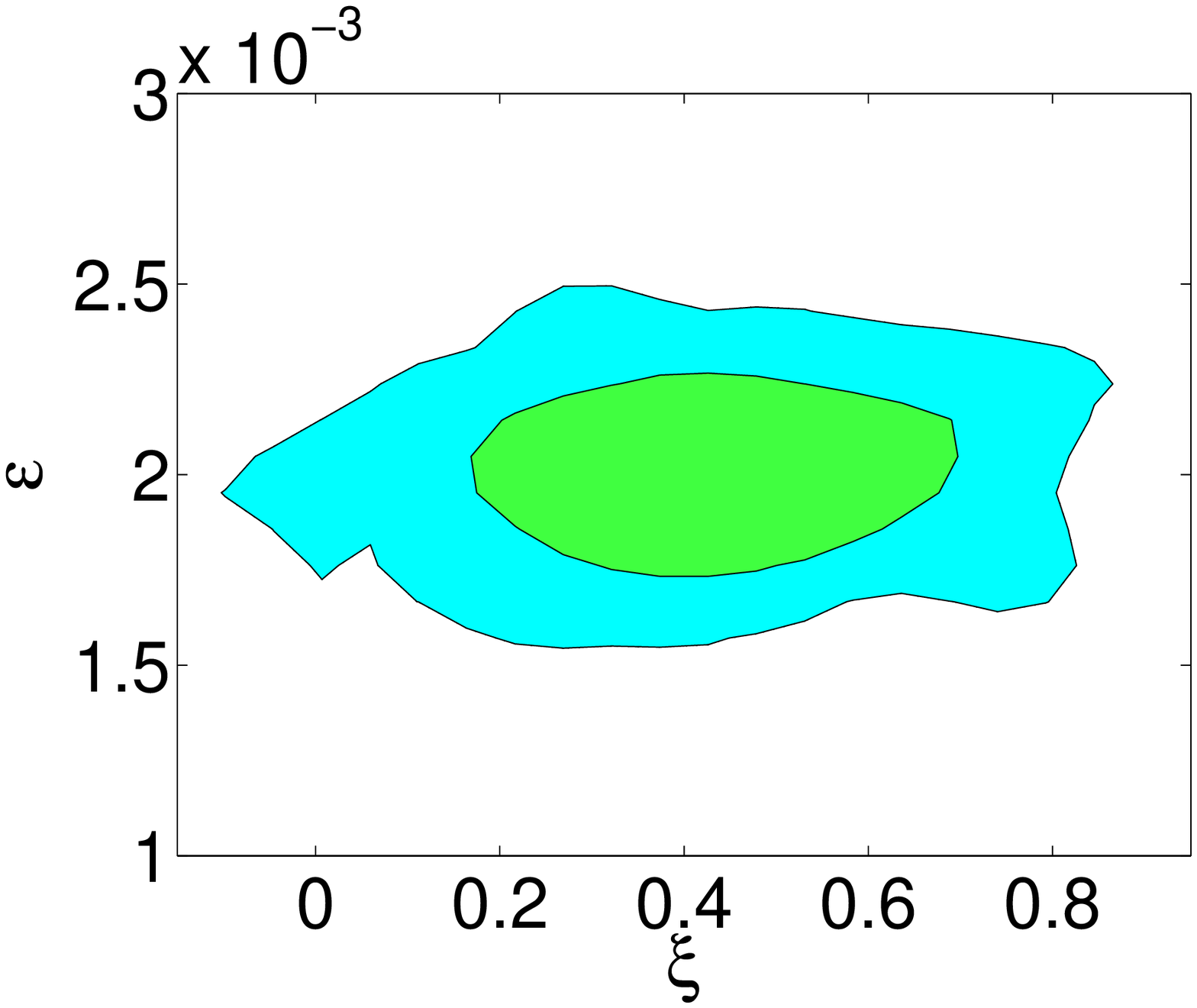,width=0.18\linewidth} \\
\end{tabular}
\caption[2-dimensional contours with $1\sigma$ and $2\sigma$ regions.]
{2-dimensional constraint of the cosmological and model parameters contours
in the flat interacting GDE model in the  BD theory with $1\sigma$ and $2\sigma$ regions. To produce these plots,
SNIa+CMB+BAO+X-ray gas mass fraction data together with the BBN constraints have been used.}\label{fig:FI}
\end{figure*}

\begin{figure*}[t]
\centering
\begin{tabular}{ccccc}
\epsfig{file=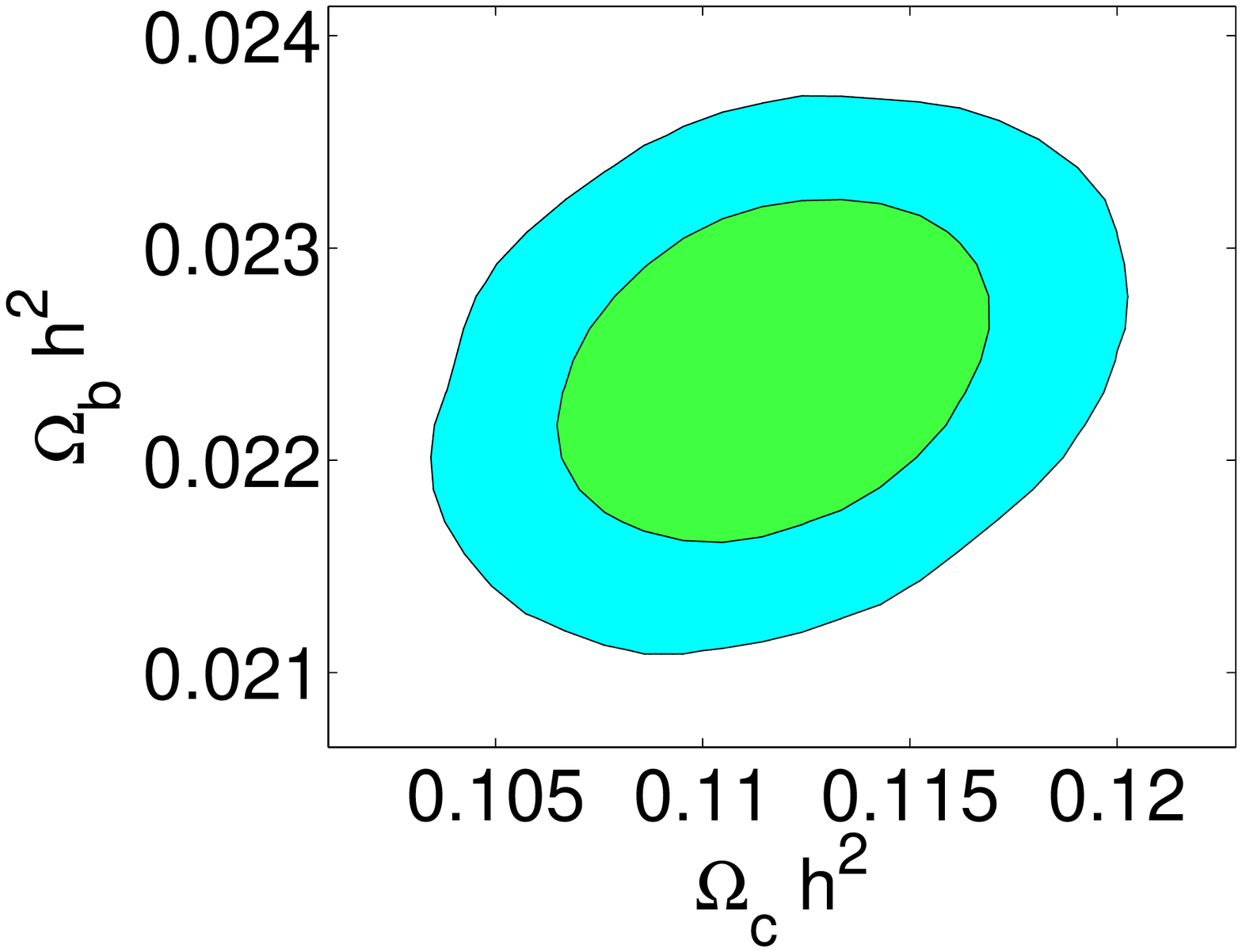,width=0.18\linewidth} & & & & \\
\epsfig{file=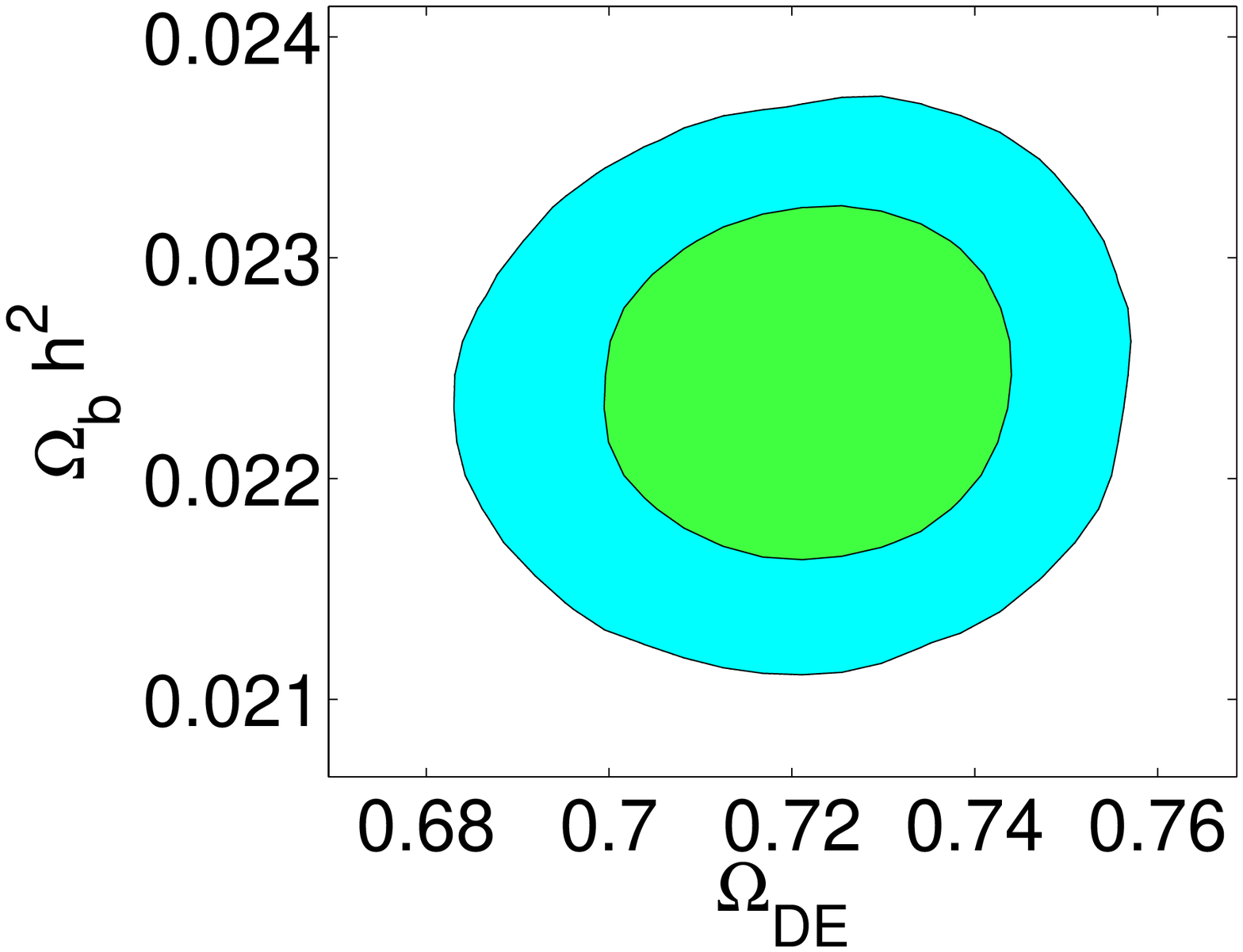,width=0.18\linewidth} & \epsfig{file=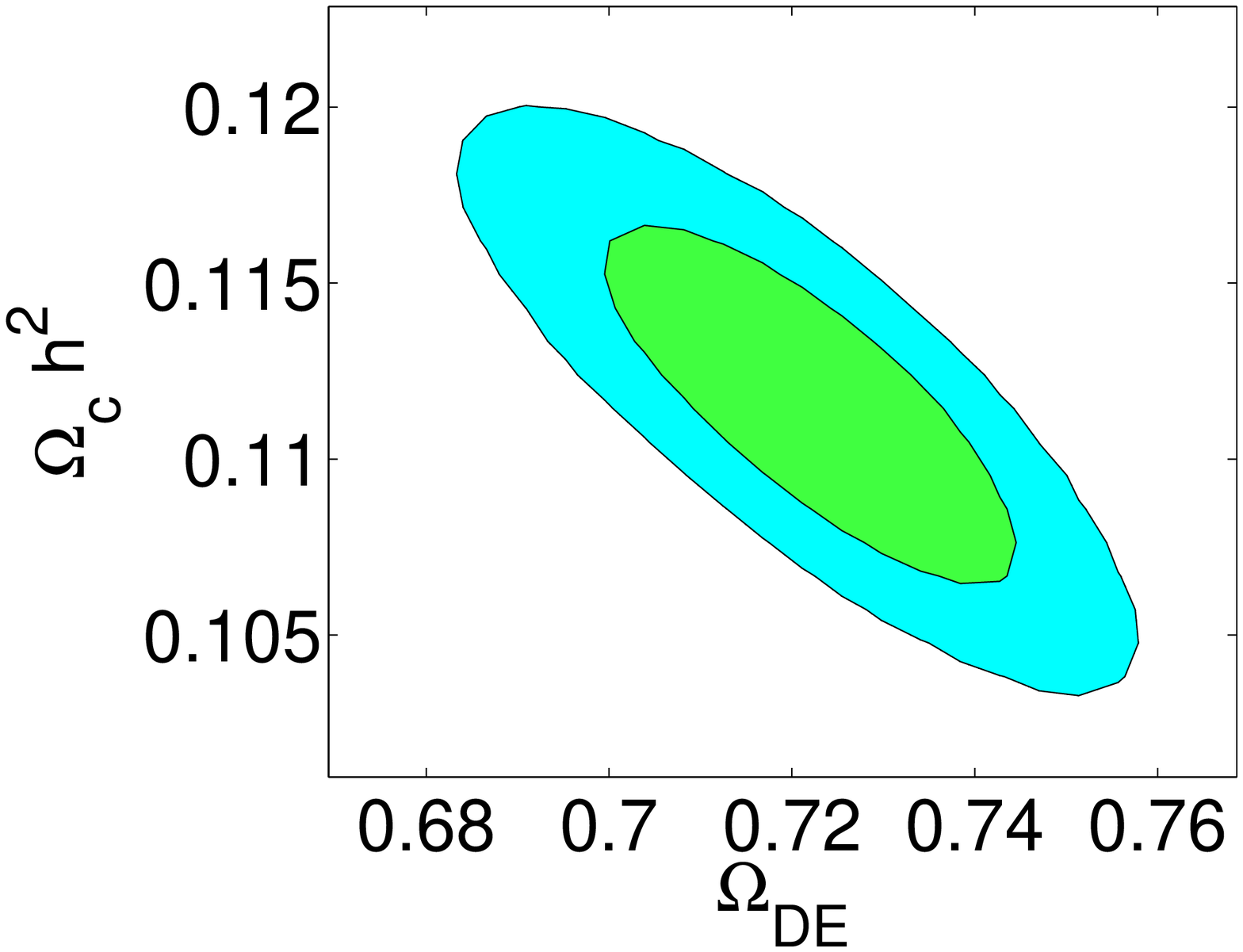,width=0.18\linewidth}& & &\\
\epsfig{file=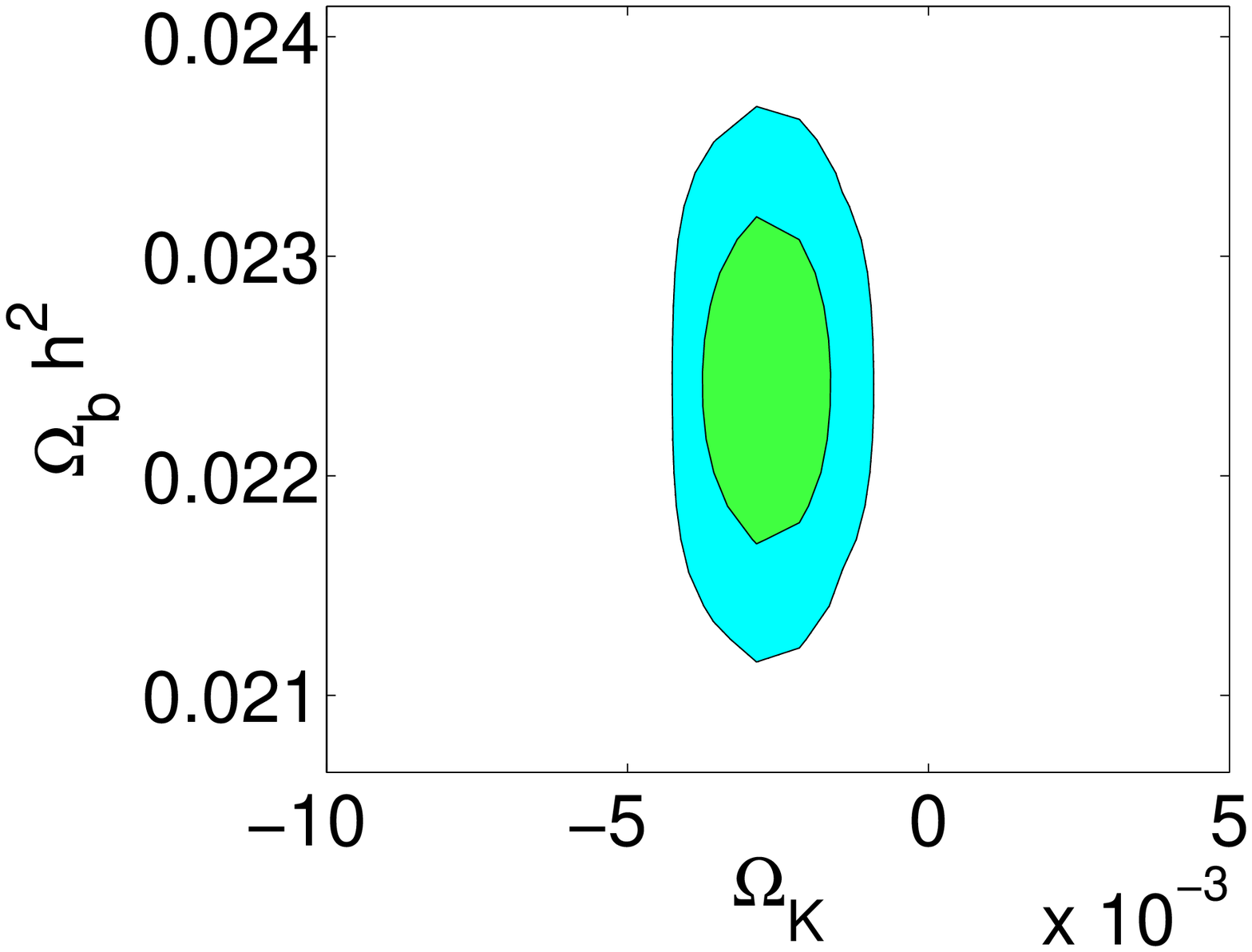,width=0.18\linewidth} & \epsfig{file=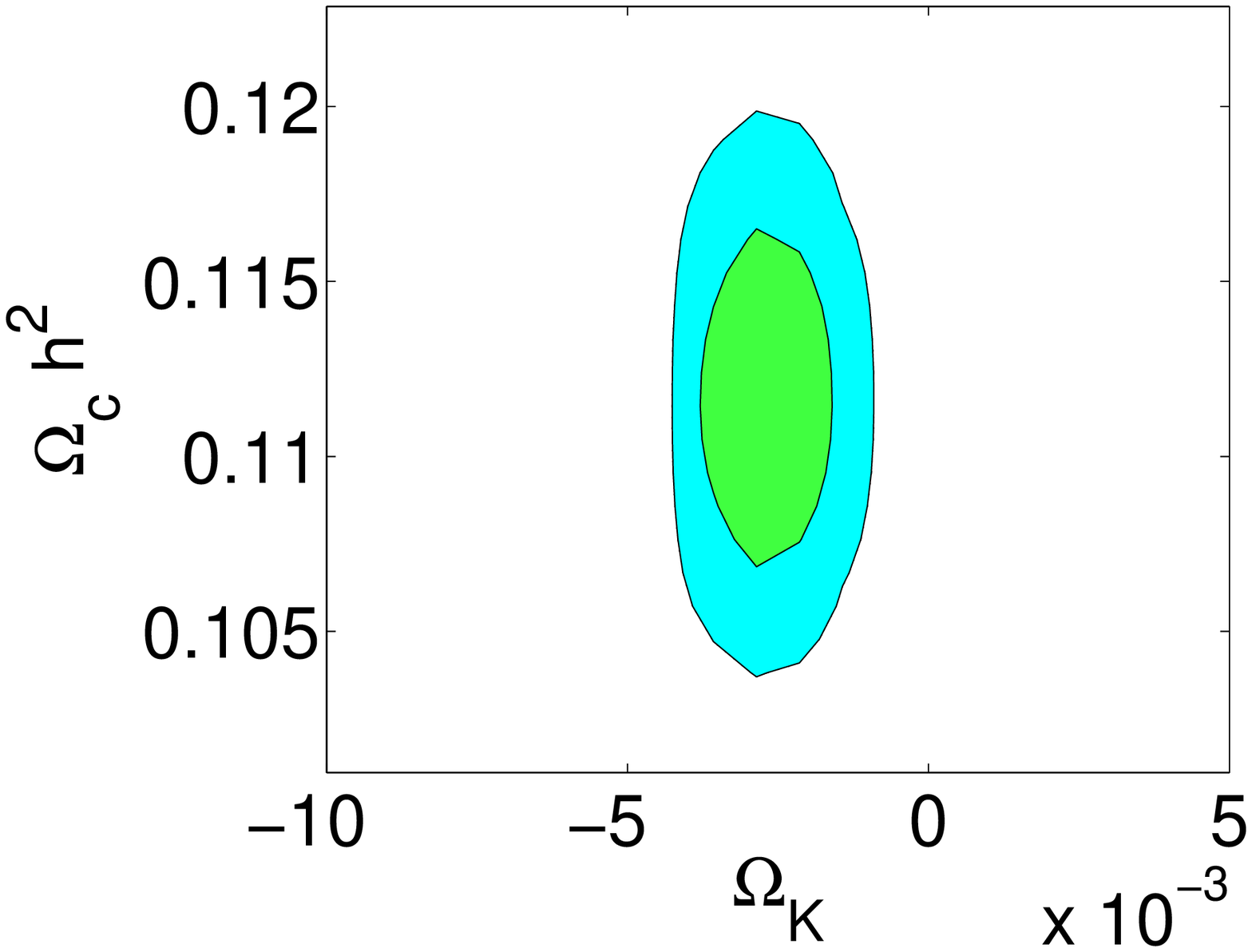,width=0.18\linewidth}&\epsfig{file=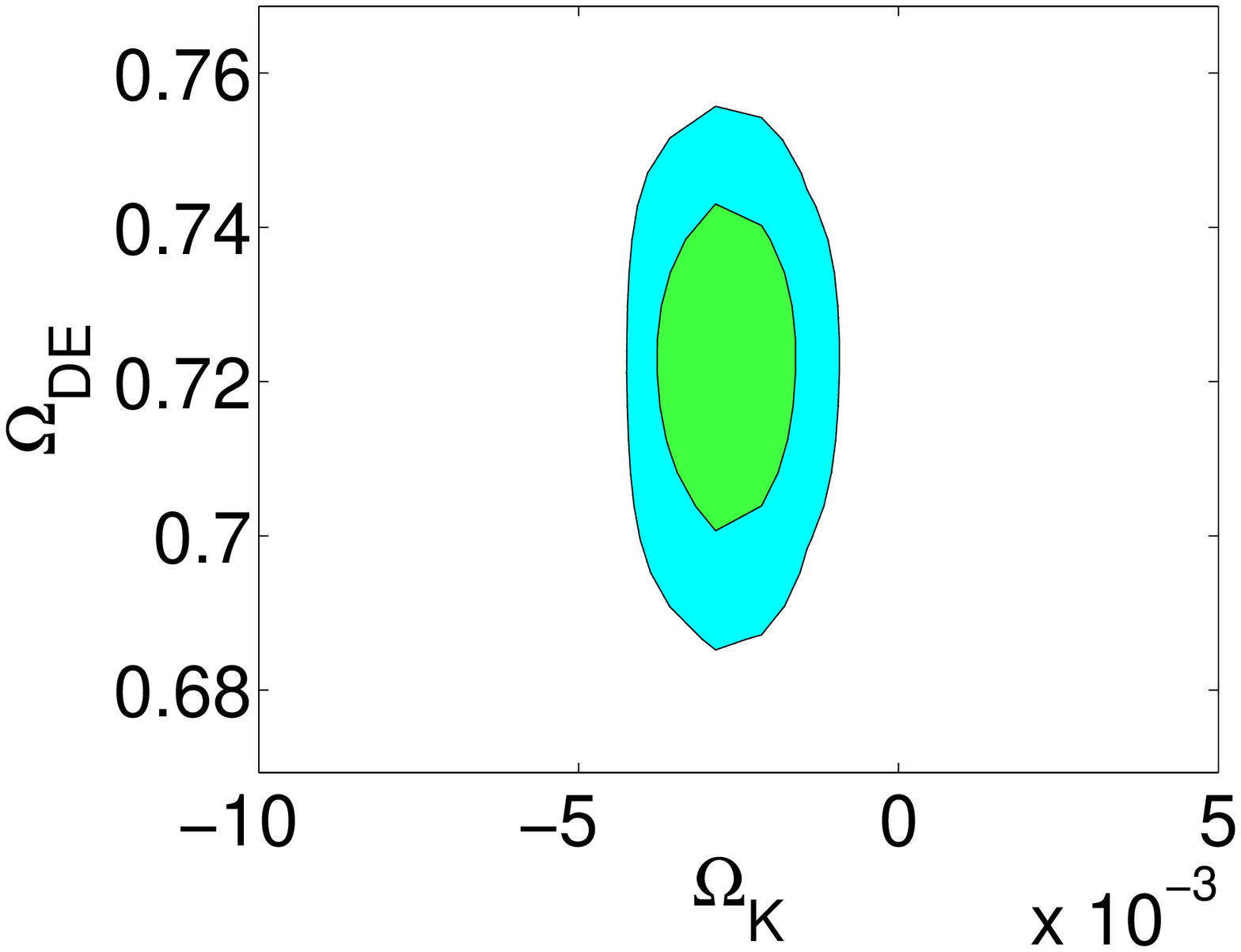,width=0.18\linewidth} & &\\
\epsfig{file=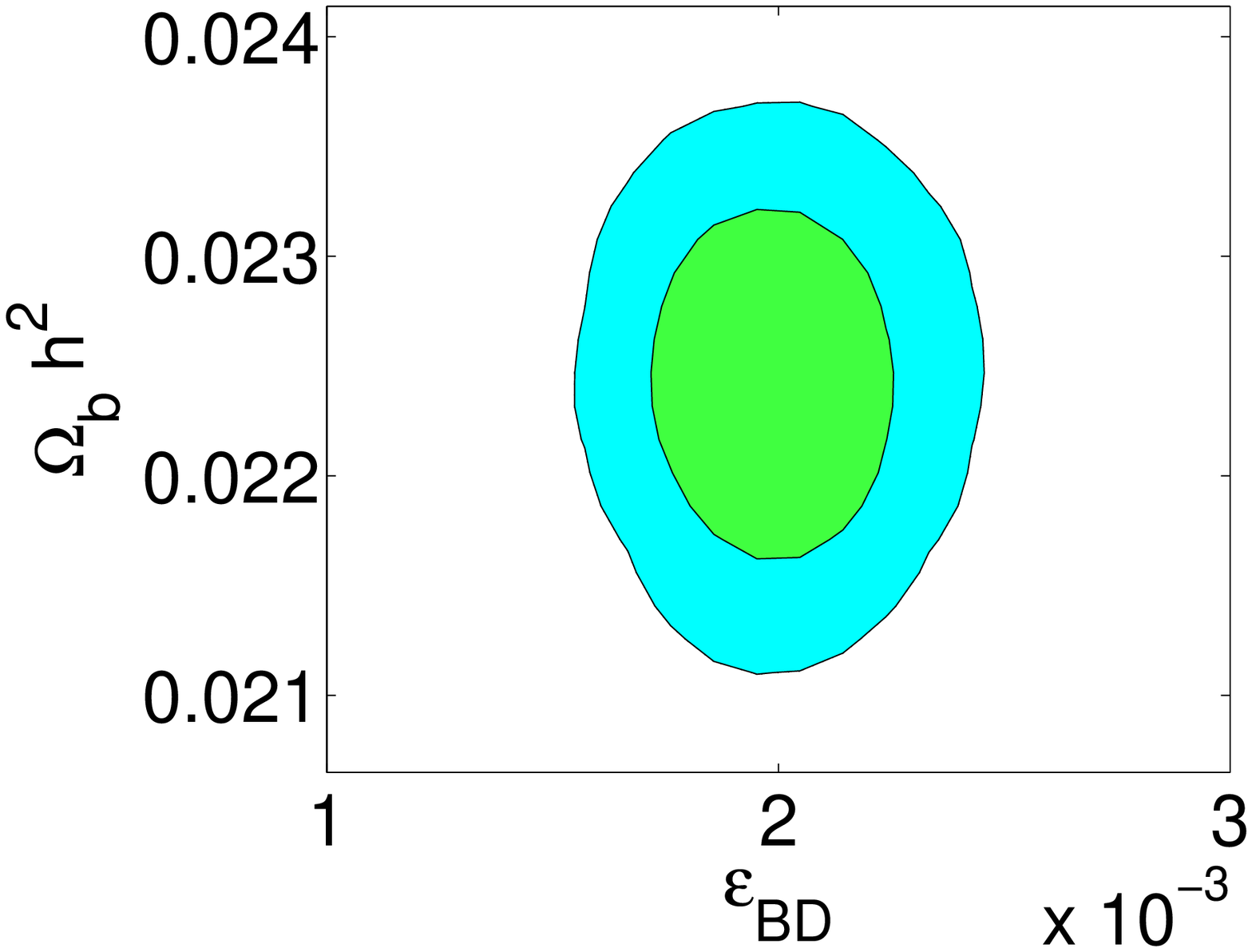,width=0.18\linewidth} &\epsfig{file=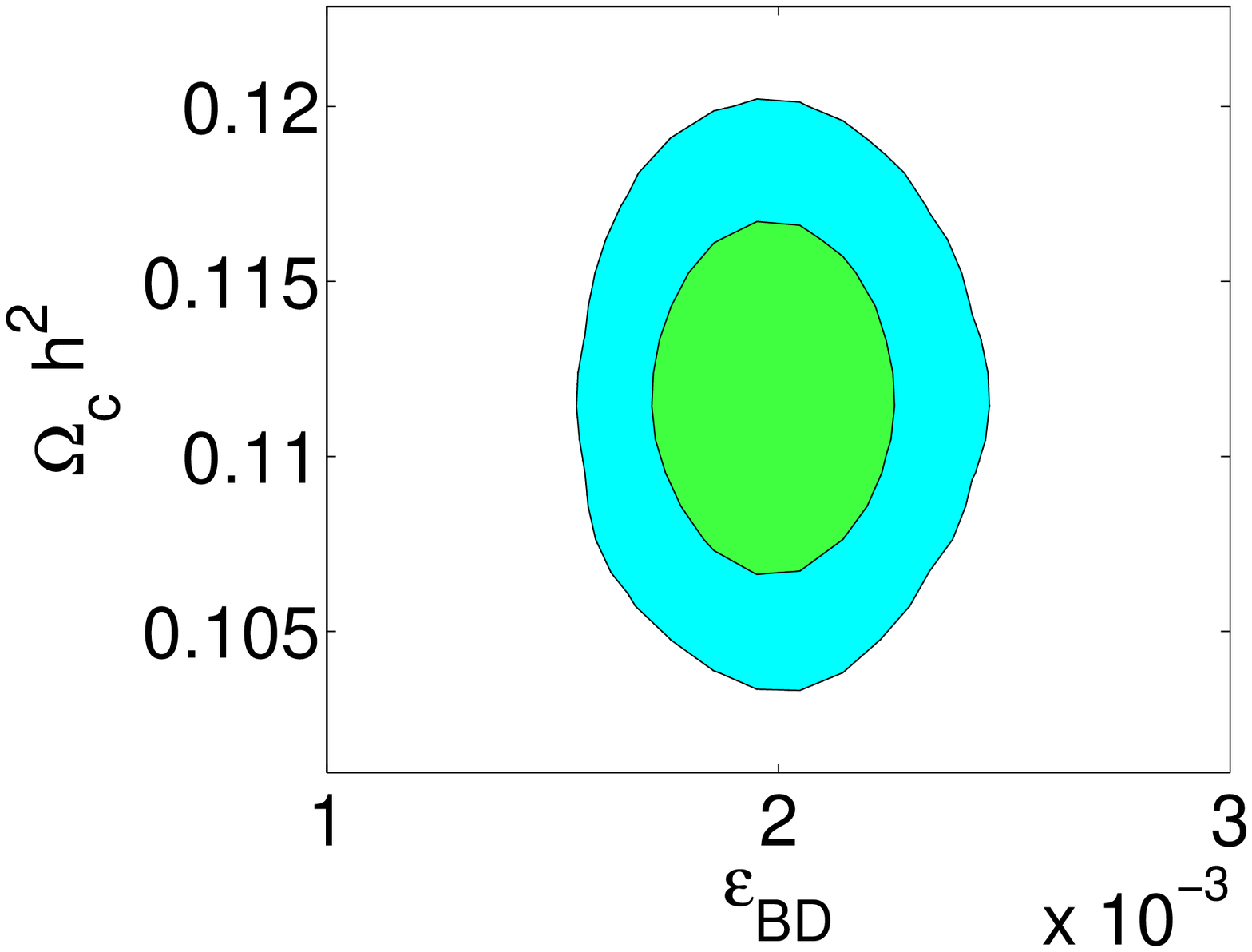,width=0.18\linewidth} & \epsfig{file=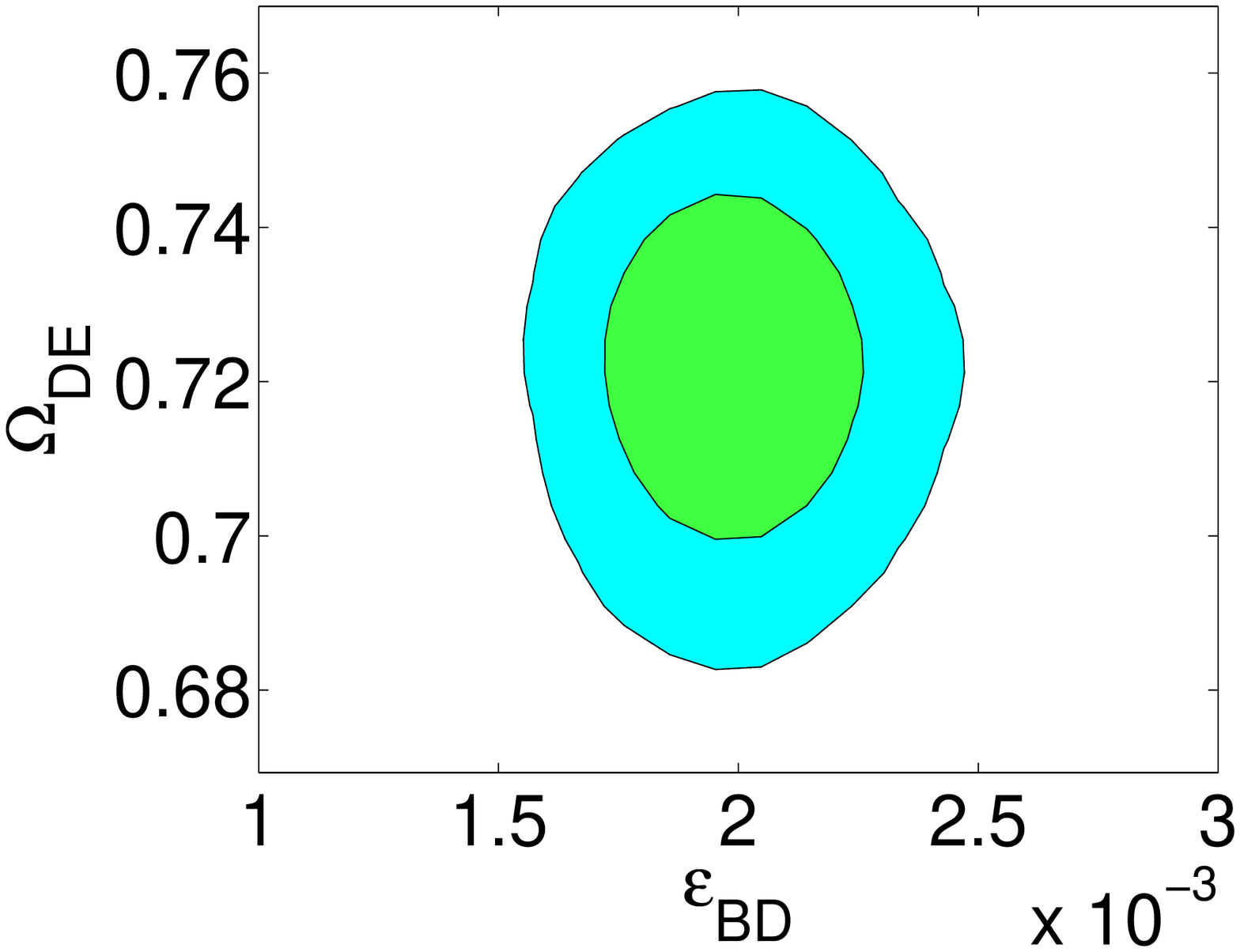,width=0.18\linewidth}& \epsfig{file=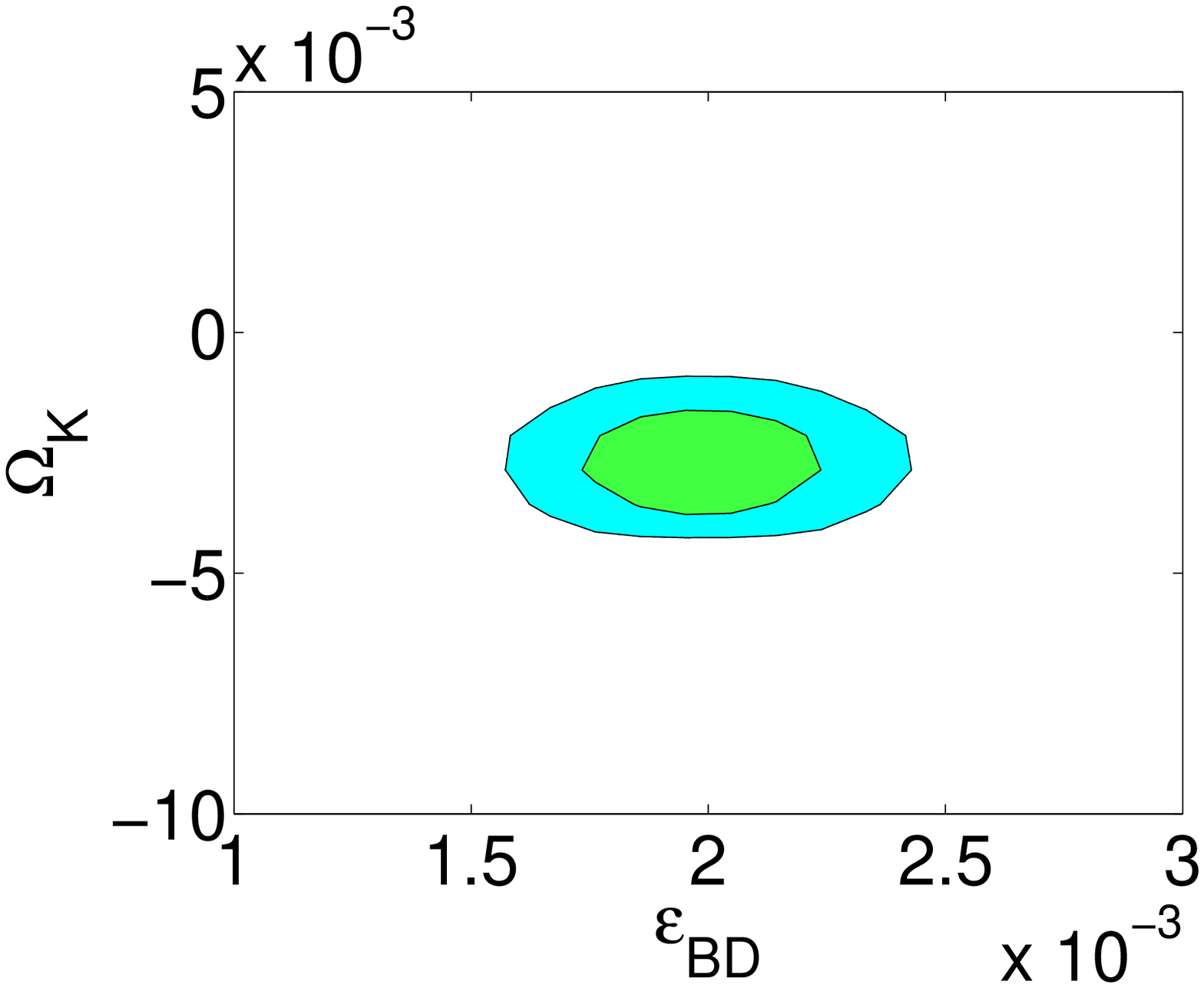,width=0.18\linewidth}&\\
\epsfig{file=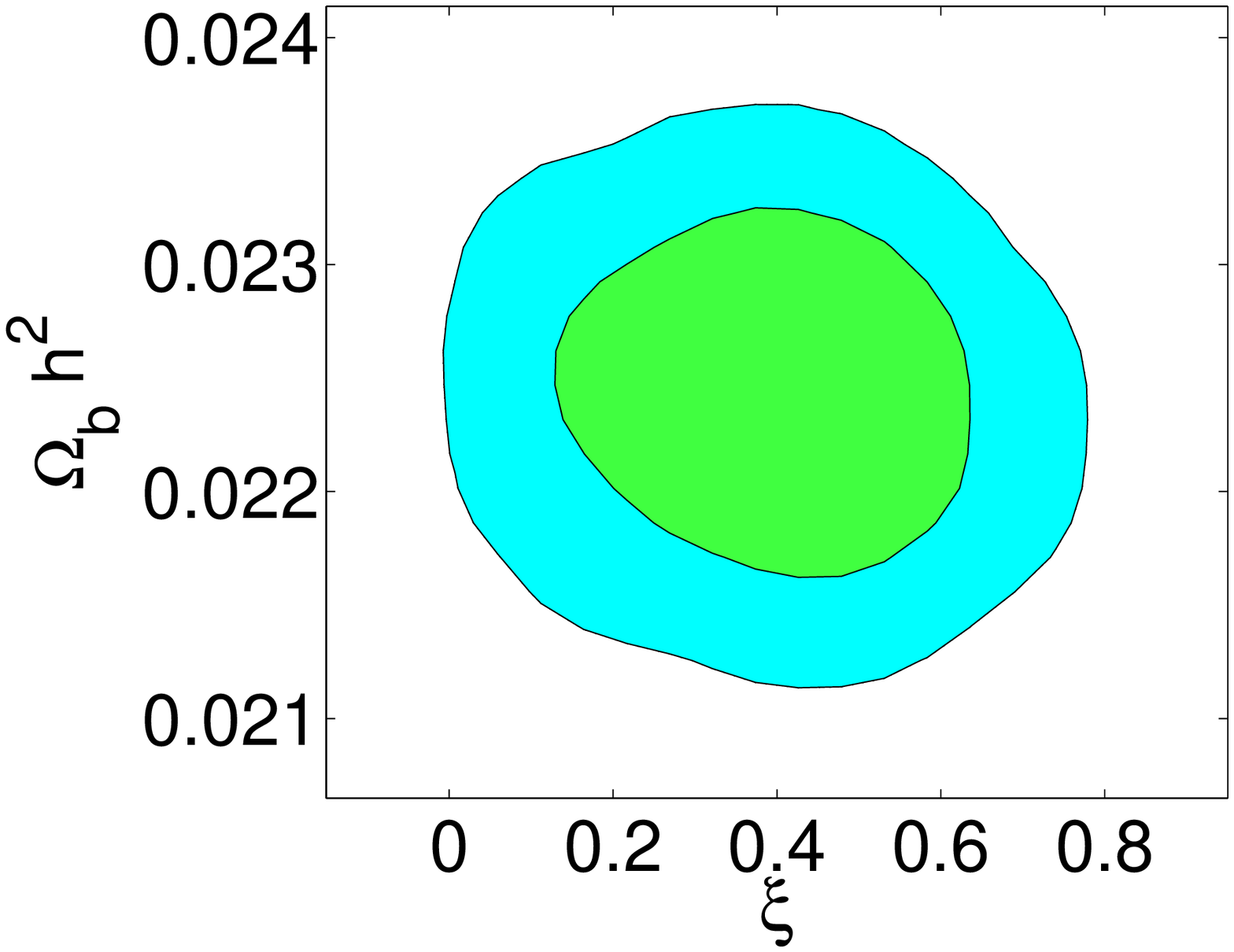,width=0.18\linewidth} &\epsfig{file=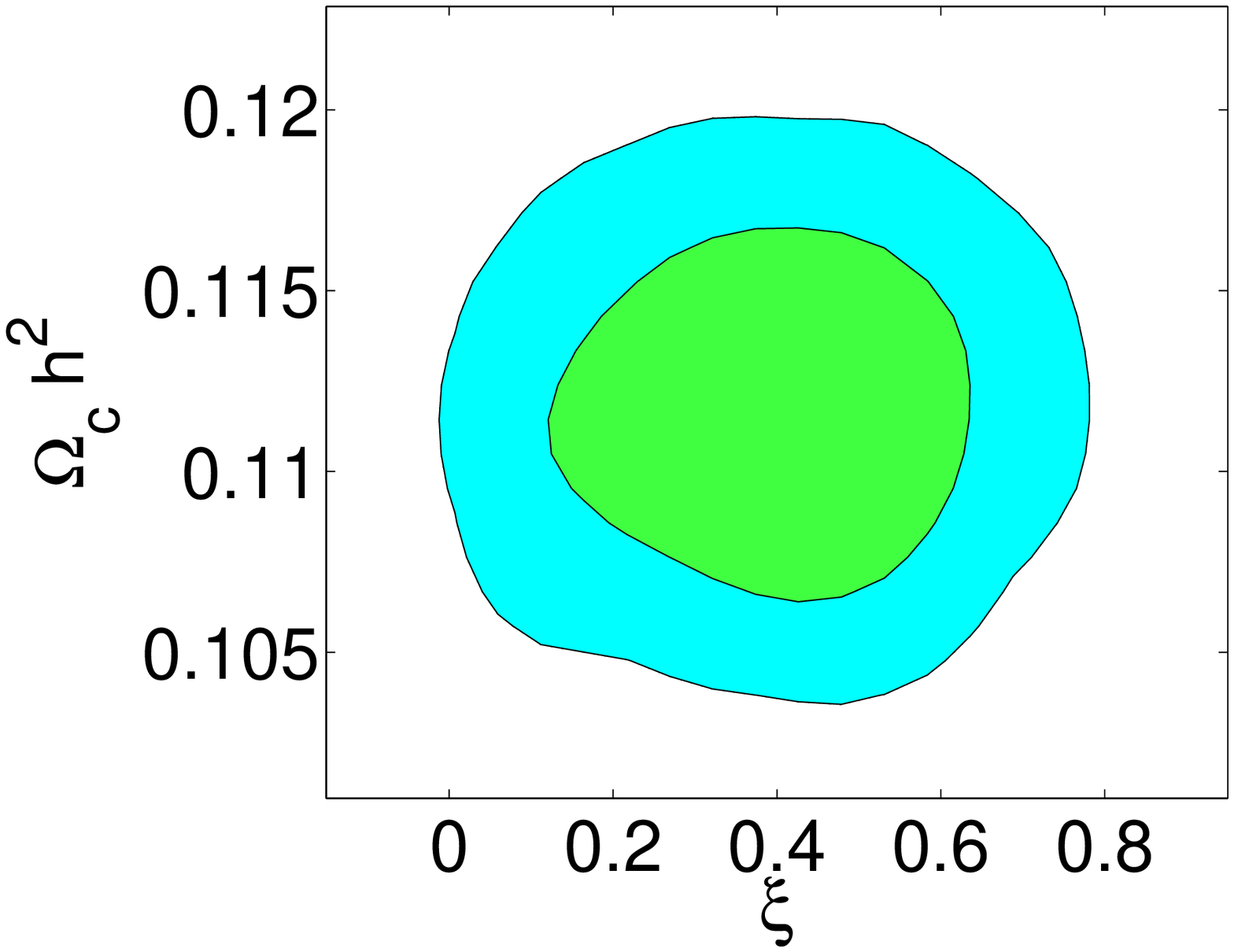,width=0.18\linewidth} & \epsfig{file=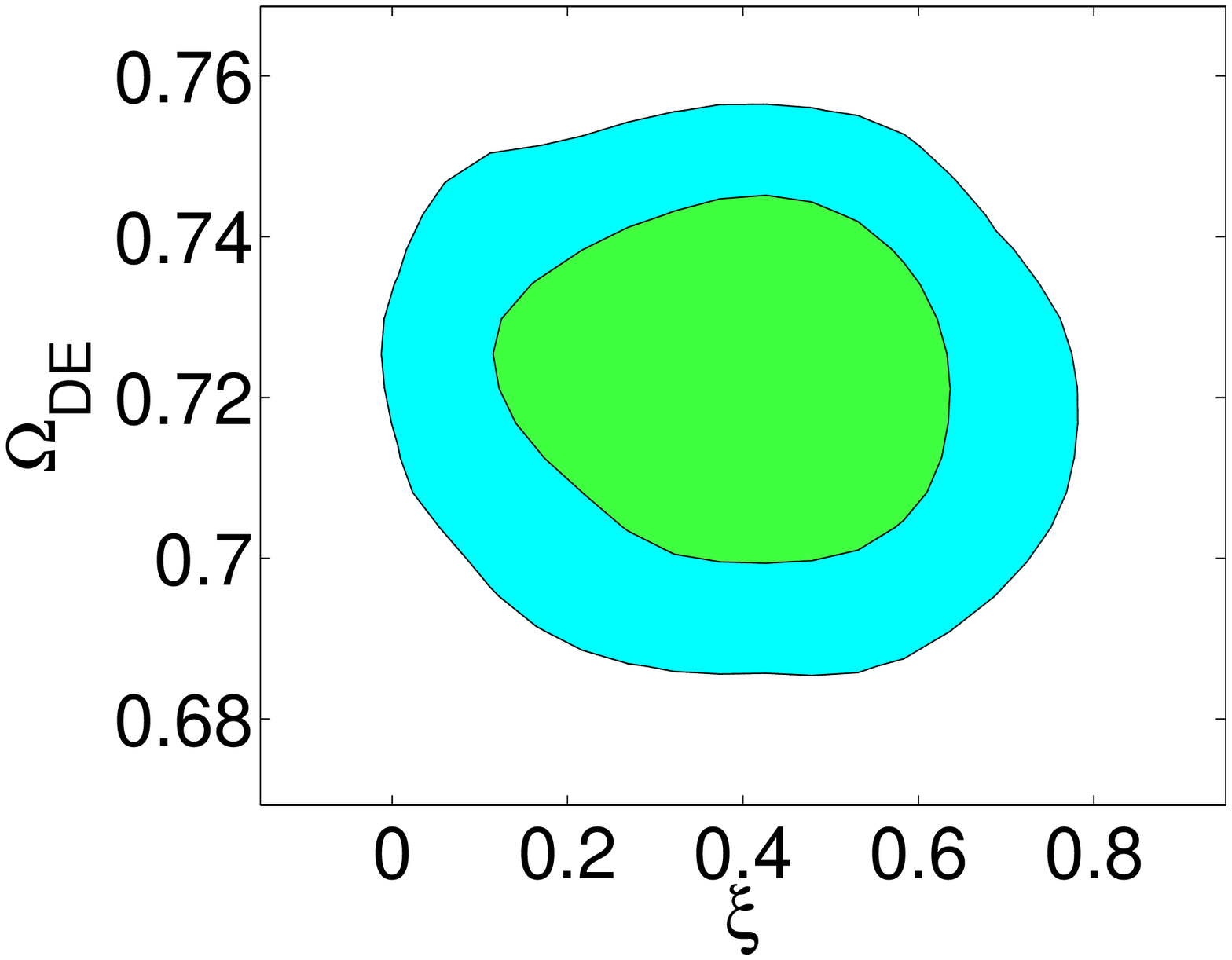,width=0.18\linewidth}& \epsfig{file=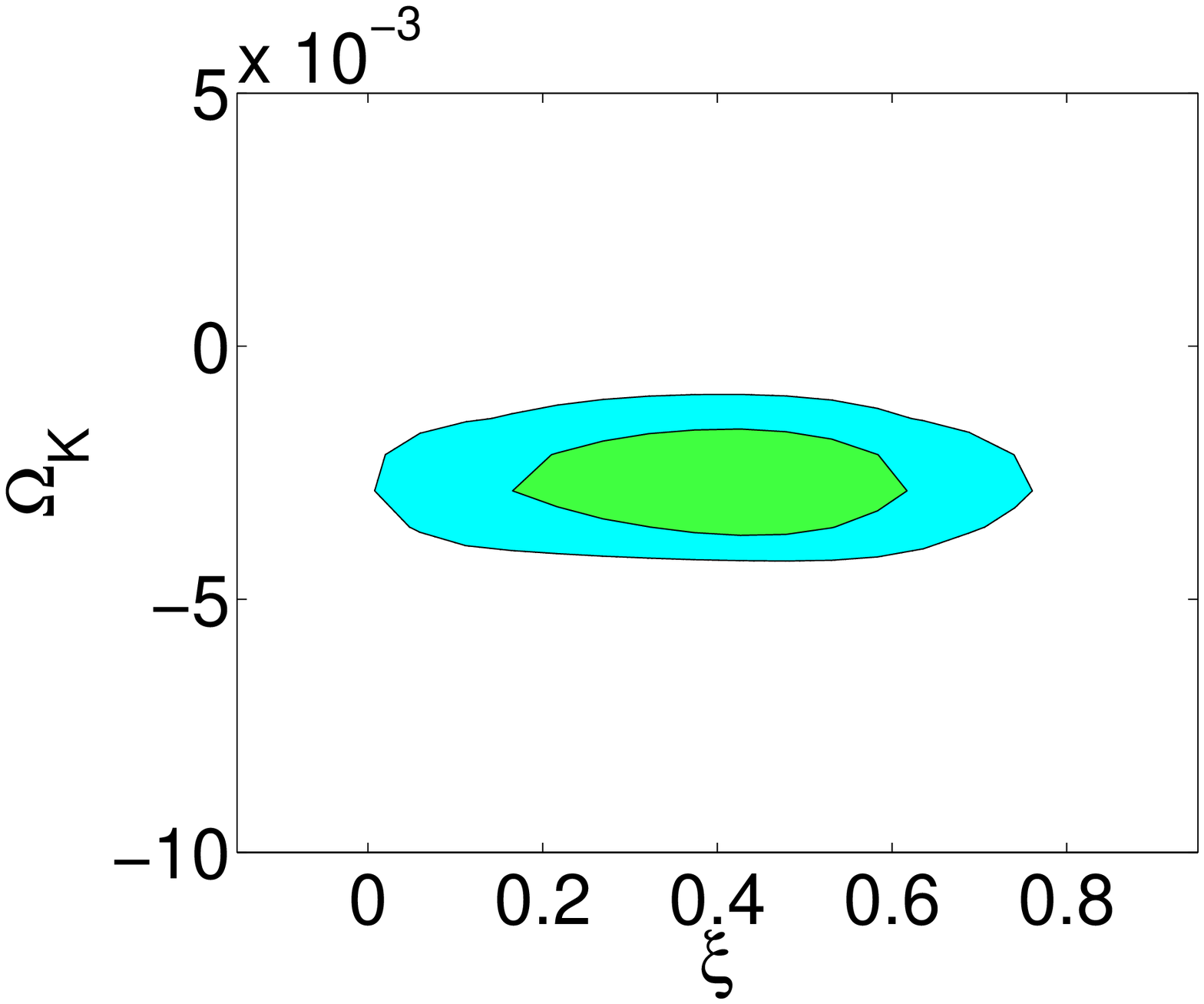,width=0.18\linewidth} &\epsfig{file=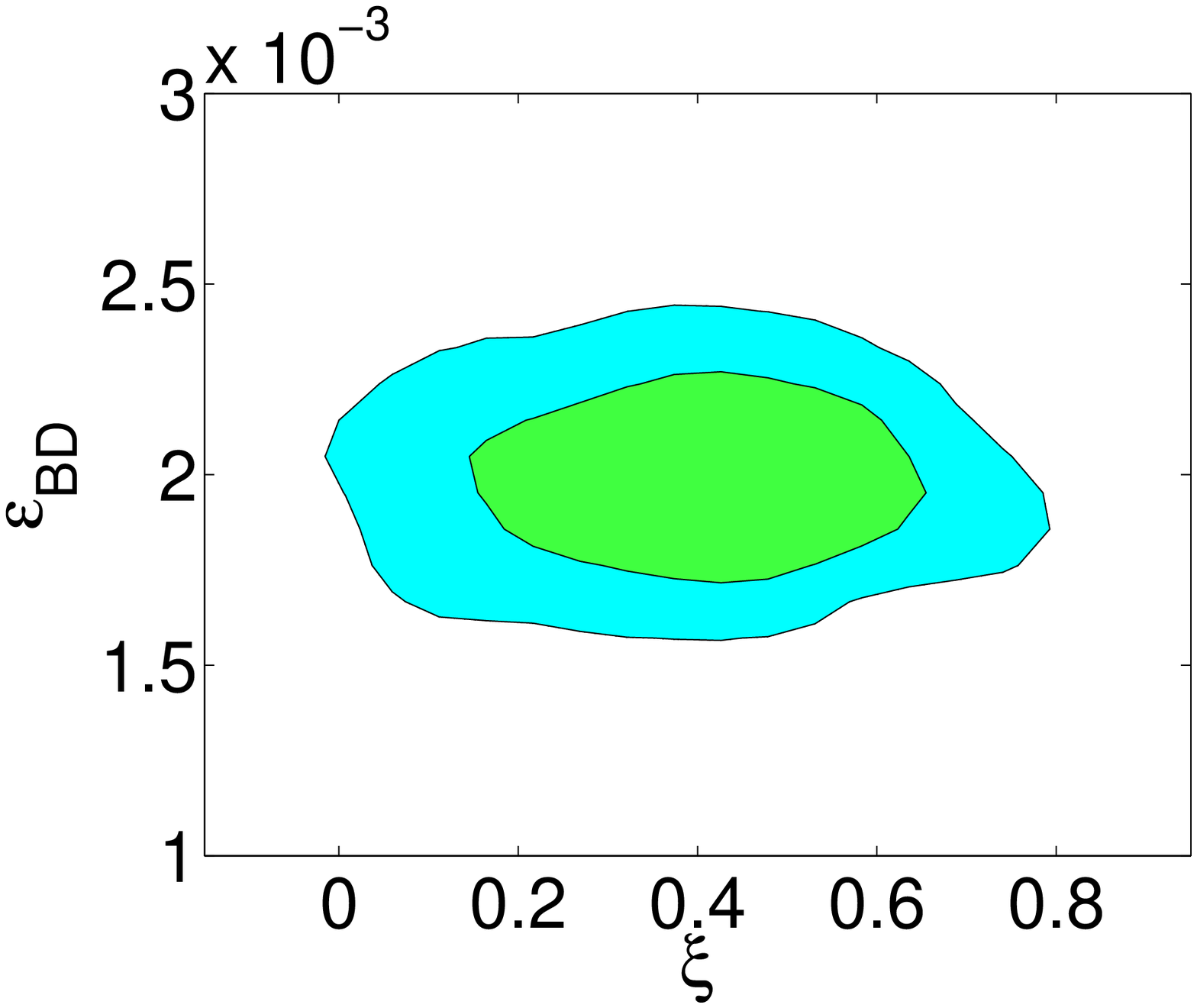,width=0.18\linewidth} \\
\end{tabular}
\caption[2-dimensional contours with $1\sigma$ and $2\sigma$ regions.]
{2-dimensional constraint of the cosmological and model parameters contours
in the non-flat interacting GDE model in the BD theory with $1\sigma$ and $2\sigma$ regions. To produce these plots,
SNIa+CMB+BAO+X-ray gas mass fraction data together with the BBN constraints have been used.}\label{fig:FI}
\end{figure*}

\section{Conclusion}\label{sec:conc}
In this paper we considered the cosmological constraints on the  parameters of
the GDE in the framework of BD theory by using a Markov Chain Monte Carlo simulation.
We used the SNIa+ CMB+ BAO+X-ray gas mass fraction data for the model fitting.
The best fit values of the cosmological
parameters in this model are compatible with the results of the $\Lambda$CDM model.  In addition, we obtained the best fit values of
parameters $\epsilon=1/\omega$ and $\xi$ where $\omega$ is the BD parameter and $\xi$ is the interacting parameter.
The best fit values of these parameters are also compatible with the results of previous constraining works.
However as we mentioned in section \ref{sec:MCMCmethod}, due to large systematic error
in  $\epsilon$ by using the supernovae data only, to constrain this parameter  with a higher precision,
we should combine the supernovae data
with other cosmological data sets as the CMB and BAO data sets.
In addition we should assert one more time that the data points parameters of the CMB and BAO data which we have used in this paper
 are the best fit values for  $\Lambda$CDM and the error estimates are also based on the $\Lambda$CDM model.
 Therefore they are not completely accurate in this application.
The numerical results can be improved in the future works by using more recent data such as nine-yaer WMAP \cite{Bennett:2012zja} or
the Planck  \cite{Ade:2013zuv} projects.

\section*{Acknowledgements}
We would  like to
thank the anonymous referee for constructive comments.
Hamzeh Alavirad would also like to thank Qader Dorosti for helpful discussions.
The work of A. Sheykhi has been supported
financially by Research Institute for Astronomy and Astrophysics
of Maragha (RIAAM), Iran.




\bibliographystyle{elsarticle-num}



\end{document}